\documentclass[11pt,a4paper]{article}
\usepackage{jcappub}
\usepackage[T1]{fontenc}
\usepackage{lmodern}
\usepackage{amsmath,amssymb,booktabs}
\usepackage{graphicx}
\usepackage{placeins}
\usepackage{needspace}
\newcommand{\dd}{\mathrm d}
\newcommand{\HH}{\mathcal H}
\newcommand{\bn}{\boldsymbol n}
\newcommand{\bv}{\boldsymbol v}
\newcommand{\Q}{\mathcal Q}
\newcommand{\Dip}{\mathcal D}
\newcommand{\PhiW}{\Phi_{\rm W}}
\graphicspath{{figures/}}
\title{Galaxy number counts in area distance space}
\author[1,2]{Chris Clarkson,}
\author[2,3]{Roy Maartens,}
\author[1]{Stefano Zazzera}
\affiliation[1]{Centre for Theoretical Physics and Astronomy, School of Physical and Chemical Sciences, Queen Mary University of London, Mile End Road, London E1 4NS, United Kingdom}
\affiliation[2]{Department of Physics \& Astronomy, University of the Western Cape, Cape Town 7535, South Africa}
\affiliation[3]{National Institute of Theoretical \& Computational Science, Cape Town 7535, South Africa}
\emailAdd{chris.clarkson@qmul.ac.uk}
\emailAdd{rmaartens@uwc.ac.za}
\emailAdd{s.zazzera@qmul.ac.uk}
\abstract{We investigate how galaxy clustering changes when area distance is used as the radial coordinate. We derive the linear relativistic number counts for a flux-selected population and compare them with redshift and luminosity distance space. Source motion does not change the beam area at a fixed emission event, so area-distance clustering has no Kaiser distortion. Lensing changes both the angular and radial positions of sources, and can enhance area-distance clustering while weakening its correlation with redshift-space clustering. Observer motion shifts area-distance shells in the opposite direction to redshift and luminosity-distance shells, allowing their kinematic dipoles to point in opposite directions. A combination of the three count fields cancels their common intrinsic density and explicit evolution and magnification biases, leaving a signal dominated by velocities nearby and optical effects at larger distances. We relate this signal to galaxy-size magnification, so that its correlations can be inferred statistically without precise distances to individual galaxies.}

\begin{document}
\maketitle
\raggedbottom
\brokenpenalty=10000
\tableofcontents
\clearpage
\section{Introduction}

The same galaxies have different apparent clustering when we locate them using redshift, luminosity distance or area distance. Peculiar velocities and light propagation perturb these observables differently, changing the radial separations assigned to the sources. In redshift space, velocity differences produce the familiar Kaiser distortion \cite{Kaiser1987,Hamilton1998}. Relativistic number counts also include gravitational potentials, lensing and source selection \cite{Yoo2009,Bonvin2011,Challinor2011,Jeong2012,Bertacca2012}, whose contributions enter the full-sky clustering statistics \cite{Tansella2018,Castorina2022,Alonso2015}.

Luminosity distance provides a radial coordinate for standard-siren clustering even when individual redshifts are unavailable \cite{Namikawa2016,Zhang2018,Namikawa2021,Libanore2021,Libanore2022}. Its relativistic number counts have been derived in refs.~\cite{Fonseca2023,FonsecaErratum2024,Balaudo2024}. Cross-correlations with galaxies and radio tracers can constrain source biases and relativistic contributions \cite{Zazzera2025,Zazzera2026}. Here we derive the counts in area distance space and compare all three radial coordinates for the same source population, so that differences in clustering arise from the observables used to locate those sources.

Area distance relates the physical cross-sectional area of a light beam at the source to its observed solid angle,
\begin{equation}
 \dd A_s=D_A^2\dd\Omega_o.
 \label{eq:area}
\end{equation}
Here $D_A$ is the usual angular diameter distance, defined from the area measured in the source screen \cite{KristianSachs1966}. For a resolved ruler it follows from an area-equivalent angular size; a single diameter can also depend on shear and the ruler's orientation \cite{Fleury2015,Schmidt2012} (see appendix~\ref{app:angular-distance}). In a spatially flat FLRW background,
\begin{equation}
 \bar D_A=ar=\frac{r}{1+z},\qquad \bar D_L=\frac r a,
 \label{eq:distances}
\end{equation}
and distance duality gives $D_L=(1+z)^2D_A$ when photons follow null geodesics and photon number is conserved \cite{Etherington1933}. The background area distance reaches a maximum $D_{A,*}=H^{-1}(z_*)$ where $r\HH=1$, with $\HH=aH$ and $c=1$. For the flat matter-plus-$\Lambda$ model of appendix~\ref{app:numerical-spectra}, $z_*=1.588$.

The leading local Newtonian terms distinguish the three choices:
\begin{align}
 \Delta_z^{\rm N}&=\delta_g-\frac1{\HH}\partial_r v_r,\nonumber\\
 \Delta_L^{\rm N}&=\delta_g-\frac{2r}{1+r\HH}\partial_r v_r,\nonumber\\
 \Delta_A^{\rm N}&=\delta_g.
 \label{eq:newtonian-counts}
\end{align}
Here $r$ is the background comoving distance, $v_r$ is the source velocity along the outward line of sight, and $\delta_g=b\delta_m^{\rm com}$. Spatial derivatives are at fixed conformal time, and the superscript N retains only the density and velocity-gradient terms. Peculiar velocity changes redshift and luminosity distance, so velocity differences between neighbouring sources distort their radial separation. At a fixed emission event, however, source motion does not change the cross-sectional area of the light beam \cite{Sachs1961}.
There is consequently no source Doppler displacement in area distance and no Kaiser term in $\Delta_A$.

The independence of beam area from source velocity holds beyond linear order. Velocities still affect the full counts through the number current and source selection, while lensing changes both the angular and radial assignment of sources. Comparing the three count fields allows us to separate these effects: a weighted combination removes the common intrinsic density and isolates the redshift perturbation at fixed area distance, with no explicit evolution or magnification bias. 
Observer motion shifts area-distance shells in the opposite direction to redshift and luminosity-distance shells, so the kinematic dipoles can also point in opposite directions.

Area-distance information can come from joint measurements of $D_L$ and $z$, through $D_A=D_L/(1+z)^2$, or from statistical estimates using galaxy sizes \cite{Alsing2015,Ye2026}. For a fixed emission event, the source Doppler terms cancel in the distance ratio. At fixed observed redshift, however, peculiar velocity changes which emission event is selected and hence the apparent size of the source. This is Doppler magnification \cite{Bonvin2008,Bacon2014,BonvinDipole2017,Andrianomena2019}. Size correlations therefore provide distance information statistically, without assigning a precise distance to each galaxy. In section~\ref{sec:applications} we use these correlations to infer area-distance clustering and discuss the resulting geometrical tests.
The radial expansion below applies away from the area-distance maximum. Appendix~\ref{app:branches} gives the treatment of the non-monotonic distance map.

\section{Number counts in area distance space}
\label{sec:counts}

Overbars denote background values. We normalise the background scale factor by $a(\bar\eta_o)=1$, so that $z=a^{-1}-1$ labels the background source position. We take $\bn$ to point from the observer to the source. The scalar-perturbed metric in Poisson gauge is
\begin{equation}
 \dd s^2=a^2[-(1+2\Psi)\dd\eta^2+(1-2\Phi)\dd\boldsymbol x^2].
 \label{eq:metric}
\end{equation}
A prime denotes $\partial_\eta$, $\HH=a'/a$, and $r=\bar\eta_o-\eta$ labels position on the background incoming ray. The total derivative along this ray is $\dd/\dd r=\partial_r-\partial_\eta$, with the observer event held fixed. We use the Weyl potential $\PhiW=(\Phi+\Psi)/2$. Local fields are evaluated at the source unless labelled $o$.

We centre the spatial coordinates on the actual observer event. In the main text we keep the observer velocity explicit and suppress the observer potential and clock terms, which are independent of direction at first order. Appendix~\ref{app:observer} restores these terms for an observer at fixed proper age.

\subsection{The source population and flux selection}

Let $N_A(\bn,D_A)\dd D_A\dd\Omega_o$ count the sources in an observed cell, with fluctuation $\Delta_A=(N_A-\bar N_A)/\bar N_A$. Their physical density is $\rho=a^{-3}n$, where $n$ is the comoving density. A bolometric flux cut $F>F_{\rm cut}$ imposes the intrinsic luminosity threshold
\begin{equation}
 L_{\rm cut}=4\pi F_{\rm cut}D_L^2
           =4\pi F_{\rm cut}(1+z)^4D_A^2.
 \label{eq:threshold}
\end{equation}
As the source distance changes, the same flux cut selects a different range of intrinsic luminosities. The selected number density therefore changes both through source evolution and through the changing threshold. We separate these effects using the evolution and magnification biases \cite{MaartensBias2021},
\begin{equation}
 b_e=\left.\frac{\partial\ln\bar n}{\partial\ln a}\right|_{L_{\rm cut}},
 \qquad
 \Q=-\left.\frac{\partial\ln\bar n}{\partial\ln L_{\rm cut}}\right|_a.
 \label{eq:biases}
\end{equation}

The derivative defining $b_e$ holds the intrinsic luminosity threshold fixed. Along the past lightcone of a flux-selected catalogue, the threshold also changes, giving
\begin{equation}
 \frac{\dd\ln\bar n_{\rm sel}}{\dd\ln a}
 =b_e+2\Q\left(1+\frac1{r\HH}\right).
 \label{eq:selected-evolution}
\end{equation}
Using this total derivative for $b_e$ in an expression that already includes the $\Q$ term would count the threshold evolution twice. 
Other flux, size or detection cuts change the selected population in different ways \cite{Jain2011,Duncan2022,Zazzera2024}; appendix~\ref{app:general-selection} treats a general scalar threshold.

We describe intrinsic clustering by a linear bias on matter-comoving hypersurfaces,
\begin{equation}
 \delta_g=b\delta_m^{\rm com},\qquad
 \delta_n^{\rm P}=\delta_g+(3-b_e)\HH V,\qquad
 \bv=\boldsymbol\nabla V.
 \label{eq:gauge}
\end{equation}
We assume that the tracers share the matter velocity. The second expression converts their density contrast from matter-comoving hypersurfaces to Poisson gauge \cite{Jeong2012}.

A distance perturbation changes the emission time selected by a fixed observed $D_A$. We must evaluate both the background density and the luminosity threshold at this shifted time. The background derivative
\begin{equation}
 \frac{\dd\bar D_A}{\dd z}=\frac{a^2}{\HH}(1-r\HH),
 \label{eq:distance-derivative}
\end{equation}
relates the distance perturbation to this time shift, giving
\begin{equation}
 \left.\frac{\delta\rho}{\bar\rho}\right|_{D_A}
 =\delta_g+(3-b_e)\HH V
 -(3-b_e)\frac{r\HH}{1-r\HH}\frac{\delta D_A}{\bar D_A}
 -\Q\left.\delta\ln L_{\rm cut}\right|_{D_A},
 \label{eq:density-remapping}
\end{equation}
where the perturbations on the right are defined at fixed background emission time. The term proportional to $\delta D_A$ accounts for the time shift. The luminosity threshold changes as well: at fixed observed $D_A$, eq.~\eqref{eq:threshold} gives
\begin{equation}
 \left.\delta\ln L_{\rm cut}\right|_{D_A}
 =4\left[\frac{\delta z}{1+z}
       -\frac{r\HH}{1-r\HH}\frac{\delta D_A}{\bar D_A}\right].
 \label{eq:selection}
\end{equation}
The bracket is the fractional redshift perturbation at fixed area distance: it includes the change in background redshift caused by shifting the emission time.

\subsection{The lightcone volume and the distance perturbation}

Number counts measure source worldlines crossing the past lightcone. The invariant count is the flux of the number current through that surface \cite{Ellis1971,Ellis1985,Challinor2011},
\begin{equation}
 \dd N=\rho D_A^2(-u_s^\mu k_\mu)|\dd\lambda_{\rm aff}|\dd\Omega_o,
 \qquad k^\mu=\frac{\dd x^\mu}{\dd\lambda_{\rm aff}},
 \label{eq:current}
\end{equation}
where $u_s^\mu$ is the source 4-velocity, $k^\mu$ is future directed and $\lambda_{\rm aff}$ is affine for the physical metric. Normalising $(-u_o^\mu k_\mu)_o=1$ gives $-u_s^\mu k_\mu=1+z$. The beam area is $D_A^2\dd\Omega_o$, and $(1+z)|\dd\lambda_{\rm aff}|$ is the radial thickness in the source frame. Writing $\dd N=N_X|\dd X|\dd\Omega_o$ gives, on a monotonic segment,
\begin{equation}
 N_X=\frac{\rho(1+z)D_A^2}
 {|\dd X/\dd\lambda_{\rm aff}|}.
 \label{eq:exact-radial-counts}
\end{equation}
This expression applies to an infinitesimal beam in any spacetime. For a general current, replace $\rho u_s^\mu$ by $J^\mu$ in eq.~\eqref{eq:current}.

At a fixed emission event and observer, changing the source 4-velocity changes the screen onto which the beam is projected. The connecting vectors change only by multiples of $k^\mu$, which is null and orthogonal to them. Their scalar products, and hence the beam area, are unchanged. Area distance is therefore independent of source velocity at that event \cite{Sachs1961}.

This geometrical argument does not require a perturbative expansion. An ideal area-distance map has no source Doppler radial displacement, the effect that produces Kaiser distortions and fingers of God in redshift space. Source motion still affects the evolution of structure, the number current through the lightcone and the selected population.

The local velocity term comes from the rate at which sources cross the lightcone. In a small region, take inertial coordinates in which the past cone is $t+r=0$ and a source moving with constant radial velocity follows $r=r_0+v_rt$. Here $r_0$ is its radius at $t=0$. At the intersection with the cone, $r_0=(1+v_r)r$. Thus a radial interval on the cone samples a factor $1+v_r$ more source worldlines than the corresponding interval at fixed time. This is the local $+v_r$ contribution carried by $-u_s^\mu k_\mu$ in eq.~\eqref{eq:current}.

For the background, eq.~\eqref{eq:current} and the distance Jacobian give the mean count density,
\begin{equation}
 \bar N_A=\frac{\bar n_{\rm sel}r^2}{a|1-r\HH|}.
 \label{eq:meanA}
\end{equation}
The transverse area alone does not determine the volume sampled by a distance bin. The factor $|1-r\HH|^{-1}$ comes from converting its radial interval to an interval along the lightcone, and would be missed by treating $D_A$ as a radial proper distance and using $D_A^2\dd D_A\dd\Omega_o$.

\Needspace{7\baselineskip}
With the observer scalar terms suppressed, the null condition at fixed emission time gives the radial displacement $\delta r=2\int_0^r\PhiW\dd\tilde r$. The angular Jacobian contains lensing and observer aberration, while the spatial metric contributes $-\Phi$ to the fractional distance perturbation. Together they give
\begin{align}
 \frac{\delta D_A}{\bar D_A}
 &=\bv_o\cdot\bn-\Phi
   +\frac2r\int_0^r\dd\tilde r\,\PhiW-\kappa,
 \label{eq:distance-perturbation}\\
 \kappa&=\frac1r\int_0^r\dd\tilde r\,
              \frac{r-\tilde r}{\tilde r}\nabla^2_\Omega\PhiW.
 \label{eq:kappa}
\end{align}
Integrated fields lie on the background ray, and $\nabla^2_\Omega$ is the unit-sphere Laplacian. The source Doppler term familiar at fixed observed redshift arises only after shifting the emission time \cite{Bonvin2008,Bacon2014}.
With $\bn$ pointing from observer to source, the redshift perturbation is
\begin{equation}
 \frac{\delta z}{1+z}
 =v_r-\bv_o\cdot\bn-\Psi
  -2\int_0^r\dd\tilde r\,\PhiW',
 \qquad v_r=\bv_s\cdot\bn.
 \label{eq:redshift-perturbation}
\end{equation}
\noindent The distance perturbation in eq.~\eqref{eq:distance-perturbation} is defined at fixed emission time, labelled by the background redshift $z=a^{-1}-1$.

At fixed observed redshift $z_{\rm obs}$, the required emission-time shift is $\delta\eta_s=[\delta z/(1+z)]/\HH$. Evaluating $\bar D_A$ at this shifted time gives $D_A(\bn,z_{\rm obs})=\bar D_A(z_{\rm obs})+\left.\delta D_A\right|_z$, where
\begin{align}
 \frac{\left.\delta D_A\right|_z}{\bar D_A}
 &=\frac{\delta D_A}{\bar D_A}
   -\frac{1-r\HH}{r\HH}\frac{\delta z}{1+z}\nonumber\\
 &=\left(1-\frac1{r\HH}\right)v_r
   +\frac{\bv_o\cdot\bn}{r\HH}-\Phi-\kappa
   +\frac2r\int_0^r\dd\tilde r\,\PhiW\nonumber\\
 &\quad+\frac{1-r\HH}{r\HH}
       \left[\Psi+2\int_0^r\dd\tilde r\,\PhiW'\right].
 \label{eq:distance-at-observed-redshift}
\end{align}
The background factors and first-order fields are evaluated at $r=r(z_{\rm obs})$. Source velocity enters through the emission-time shift required to keep the observed redshift fixed \cite{Bonvin2006,HuiGreene2006}.

Conversely, holding the observed area distance fixed shifts the emission time by $\delta\eta_s=r(\delta D_A/\bar D_A)/(1-r\HH)$. The resulting fractional redshift perturbation is
\begin{equation}
 \left.\frac{\delta z}{1+z}\right|_{D_A}
 =\frac{\delta z}{1+z}
  -\frac{r\HH}{1-r\HH}\frac{\delta D_A}{\bar D_A}
 =-\frac{r\HH}{1-r\HH}
       \frac{\left.\delta D_A\right|_z}{\bar D_A}.
 \label{eq:fixedDA-redshift}
\end{equation}
We use eq.~\eqref{eq:distance-at-observed-redshift} for magnification and distance--redshift diagrams at fixed observed redshift, and eq.~\eqref{eq:fixedDA-redshift} for redshift fluctuations at fixed area distance. The area-distance count derivation below starts from the fixed-emission-time perturbation~\eqref{eq:distance-perturbation} and includes the emission-time shift explicitly.

Expanding the current in eq.~\eqref{eq:current}, shifting the source time and inserting the flux selection gives
\begin{align}
 \Delta_A={}&\delta_g+(3-b_e)\HH V-2\Phi+\Psi+v_r-2\kappa
 +\frac4r\int_0^r\dd\tilde r\,\PhiW \nonumber\\
 &+2\bv_o\cdot\bn
 -\frac{r}{1-r\HH}\frac{\dd}{\dd r}
       \left(\frac{\delta D_A}{\bar D_A}\right)
 -\beta_A\frac{\delta D_A}{\bar D_A}
 -4\Q\frac{\delta z}{1+z},
 \label{eq:countsA}\\
 \beta_A={}&\frac{3-2r\HH-r^2\HH'}{(1-r\HH)^2}
             -\frac{r\HH}{1-r\HH}(b_e+4\Q).
 \label{eq:betaA}
\end{align}
The terms proportional to $\delta D_A$ describe how the fixed-distance cell moves along the lightcone. Its radial deformation gives the derivative term, while $\beta_A$ combines the changes in background density, volume and luminosity threshold.

Substituting the distance and redshift perturbations gives
\begin{align}
 \Delta_A={}&\delta_g+(3-b_e)\HH V+(1-4\Q)v_r
             +(2+4\Q-\beta_A)\bv_o\cdot\bn \nonumber\\
 &+\left[-2+\beta_A-\frac1{1-r\HH}\right]\Phi
  +\left[1+4\Q-\frac1{1-r\HH}\right]\Psi
  +\frac r{1-r\HH}(\partial_r\Phi-\Phi') \nonumber\\
 &+\frac2r\left[2-\beta_A+\frac1{1-r\HH}\right]
                    \int_0^r\dd\tilde r\,\PhiW
   +8\Q\int_0^r\dd\tilde r\,\PhiW' \nonumber\\
 &+\frac1r\int_0^r\dd\tilde r\,
       \left[(\beta_A-2)\frac{r-\tilde r}{\tilde r}
                          +\frac1{1-r\HH}\right]\nabla^2_\Omega\PhiW.
 \label{eq:expanded-counts}
\end{align}
The omitted observer potential and clock terms contribute only to the monopole at first order. Equations~\eqref{eq:observer-distance}--\eqref{eq:observer-counts} in appendix~\ref{app:observer} give their additions to the distance, redshift and count perturbations. No source Euler equation has been used in eq.~\eqref{eq:expanded-counts}; the radial potential gradient remains explicit.

Since $\delta D_A/\bar D_A$ at fixed emission time contains no source velocity, differentiating it produces no source-velocity gradient. The remaining local velocity coefficient is $1-4\Q$, with $+1$ from the current and $-4\Q$ from flux selection. The distance derivative also produces a radial potential gradient. Combining that gradient with the velocity terms using the source Euler equation changes their coefficient, as we show below.

\section{Comparison with luminosity distance and redshift}
\label{sec:comparison}

We use the same bolometric flux-selected population in all three spaces, with redshift $z$ labelling the background source position. Distance duality gives
\begin{equation}
 \frac{\delta D_L}{\bar D_L}
 =\frac{\delta D_A}{\bar D_A}+2\frac{\delta z}{1+z}.
 \label{eq:duality-perturbation}
\end{equation}
The analogous luminosity-distance coefficient is
\begin{equation}
 \beta_L=\frac{3+2r\HH+r^2\HH'}{(1+r\HH)^2}
            -\frac{r\HH}{1+r\HH}b_e.
 \label{eq:betaL}
\end{equation}
At fixed $D_L$, the intrinsic threshold $4\pi F_{\rm cut}D_L^2$ is also fixed. The luminosity-distance coefficient consequently has no explicit $\Q$ dependence. This simplification refers to the bolometric flux cut; appendix~\ref{app:general-selection} gives the selection contribution for a general scalar threshold.

\subsection{Source velocities and lensing}

The velocity coefficients depend on the source dynamics. Geodesic sources obey the Euler equation,
\begin{equation}
 v_r'+\HH v_r+\partial_r\Psi=0.
 \label{eq:euler}
\end{equation}
Setting $\Phi=\Psi$, we can write the local velocity and radial potential-gradient terms as follows. The first line of each expression keeps the gradient explicit; the second uses the Euler equation:
\begin{align}
 \left.\Delta_A\right|_{v_r,\partial_r\Psi}
 &=(1-4\Q)v_r+\frac r{1-r\HH}\partial_r\Psi
 \nonumber\\
 &\overset{\mathrm{Euler}}{=}
 \left[1-4\Q-\frac{r\HH}{1-r\HH}\right]v_r
       -\frac r{1-r\HH}v_r',
 \label{eq:velocityA-before-after}\\[4pt]
 \left.\Delta_L\right|_{v_r,\partial_r\Psi}
 &=(1-2\beta_L)v_r
   +\frac r{1+r\HH}\left(3\partial_r\Psi+2v_r'-2\partial_r v_r\right)
 \nonumber\\
 &\overset{\mathrm{Euler}}{=}
 \left[1-2\beta_L-\frac{3r\HH}{1+r\HH}\right]v_r
       -\frac r{1+r\HH}v_r'
       -\frac{2r}{1+r\HH}\partial_r v_r,
 \label{eq:velocityL-before-after}\\[4pt]
 \left.\Delta_z\right|_{v_r,\partial_r\Psi}
 &=\left[1-\frac{\HH'}{\HH^2}-\frac{2-2\Q}{r\HH}-2\Q+b_e\right]v_r
       +\frac1\HH\left(v_r'+\partial_r\Psi-\partial_r v_r\right)
 \nonumber\\
 &\overset{\mathrm{Euler}}{=}
 \left[-\frac{\HH'}{\HH^2}-\frac{2-2\Q}{r\HH}-2\Q+b_e\right]v_r
       -\frac1\HH\partial_r v_r.
 \label{eq:velocityz-before-after}
\end{align}
The area-distance expression has a local $+v_r$ from the number current and $-4\Q v_r$ from the bolometric flux cut. Its potential gradient comes from the distance Jacobian. The Euler equation rewrites this gradient in terms of $v_r$ and its time derivative.

For tests of free fall, we leave $v_r'+\HH v_r+\partial_r\Psi$ unconstrained \cite{Bonvin2014,BonvinFleury2018}. It enters area-distance, luminosity-distance and redshift counts with coefficients $r/(1-r\HH)$, $3r/(1+r\HH)$ and $1/\HH$, respectively. A difference between the metric potentials gives a separate contribution: allowing $\Phi\ne\Psi$ adds $r\partial_r(\Phi-\Psi)/(1\mp r\HH)$, with the minus sign for area distance and the plus sign for luminosity distance. Keeping both contributions is needed for tests of gravity that allow departures from geodesic motion or unequal metric potentials.

\begin{figure}[t]
 \centering\includegraphics[width=\textwidth]{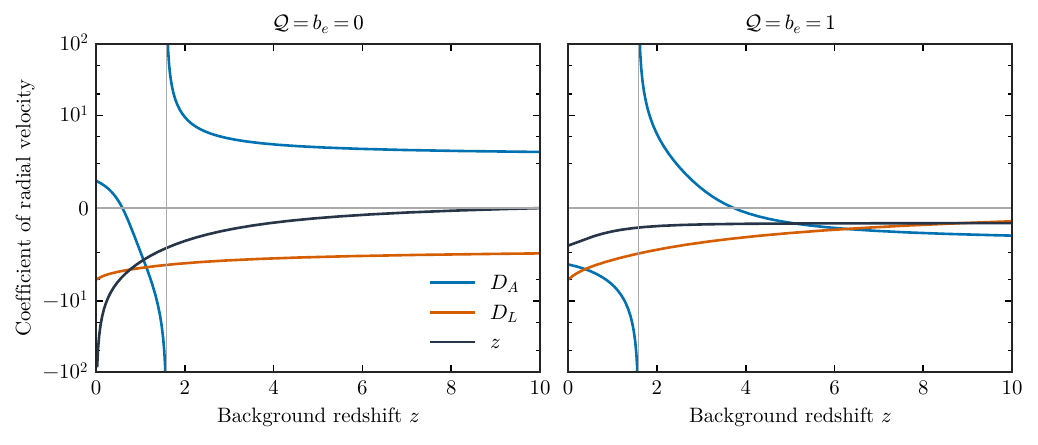}
 \caption{Local velocity coefficients in eqs.~\eqref{eq:velocityA}--\eqref{eq:velocityz}, after the same Euler and growing-mode reduction, for $\Q=b_e=0$ and $1$. The vertical axes use an inverse-hyperbolic-sine (asinh) scale, linear near zero and logarithmic at large magnitude. The area-distance coefficient diverges at the distance maximum (grey line), where $1-r\HH=0$ and the pointwise radial expansion fails.}
 \label{fig:velocity}
\end{figure}

For the numerical examples we further assume the scale-independent growing mode of a flat matter-plus-$\Lambda$ model. With linear growth factor $D_+$, $f=\dd\ln D_+/\dd\ln a$ and $\Omega_m=\Omega_m(z)$, the growth equation gives $v_r'=[3\Omega_m/(2f)-1]\HH v_r$. The velocity terms then become
\begin{align}
 \left.\Delta_A\right|_{v_r}
 &=\left[1-4\Q-\frac{3\Omega_m}{2f}\frac{r\HH}{1-r\HH}\right]v_r,
 \label{eq:velocityA}\\
 \left.\Delta_L\right|_{v_r}
 &=\left[1-2\beta_L-\left(2+\frac{3\Omega_m}{2f}\right)
                 \frac{r\HH}{1+r\HH}\right]v_r
    -\frac{2r}{1+r\HH}\partial_r v_r,
 \label{eq:velocityL}\\
 \left.\Delta_z\right|_{v_r}
 &=\left[-\frac{\HH'}{\HH^2}-\frac{2-2\Q}{r\HH}-2\Q+b_e\right]v_r
    -\frac1\HH\partial_r v_r.
 \label{eq:velocityz}
\end{align}
The coefficients multiplying $v_r$ are shown in figure~\ref{fig:velocity}. The slicing term $(3-b_e)\HH V$ remains separately in the full counts.

\begin{figure}[t]
 \centering\includegraphics[width=\textwidth]{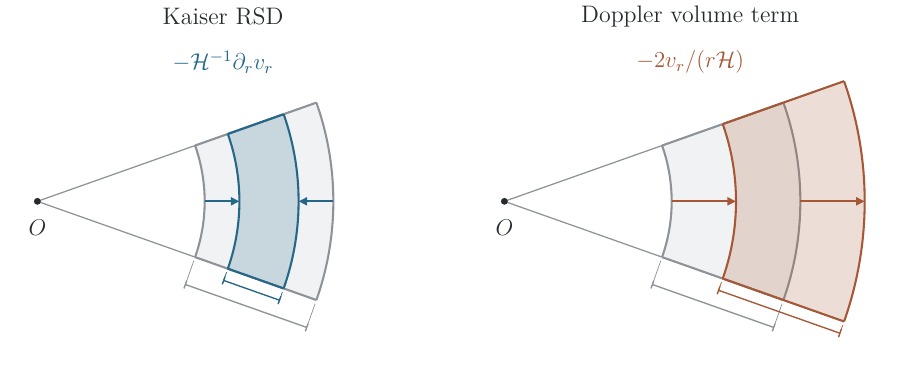}
 \caption{Radial and transverse volume changes in redshift space. Grey and coloured shells contain the same galaxies at their true and inferred positions, with $r\mapsto r+v_r/\HH$ and the observer at rest. Left: a converging velocity field compresses the inferred radial depth, giving the Kaiser term $-\partial_r v_r/\HH$. Right: a constant displacement $v_r/\HH$ moves both boundaries outwards equally, preserving their separation and increasing the transverse area. The fractional area change is $2v_r/(r\HH)$, so the inferred density decreases. The brackets mark radial depths. We hold $\HH$ constant across a narrow, nearby bin and exaggerate the displacements; eq.~\eqref{eq:velocityz} gives the full velocity coefficient.}
 \label{fig:rsd-doppler-geometry}
\end{figure}

The radial and transverse volume changes can be seen directly in the nearby redshift-space map $r\mapsto r+v_r/\HH$. Holding $\HH$ constant across the bin gives the volume Jacobian $1+\partial_r v_r/\HH+2v_r/(r\HH)$ at first order. A constant $v_r/\HH$ leaves the radial depth unchanged, but increases the transverse area by $2v_r/(r\HH)$; the $1/r$ comes from the fractional change in radius. Variation of $v_r/\HH$ across the shell changes its radial depth as well. See figure~\ref{fig:rsd-doppler-geometry} for the two geometrical contributions.

\begin{figure}[t]
 \centering\includegraphics[width=\textwidth]{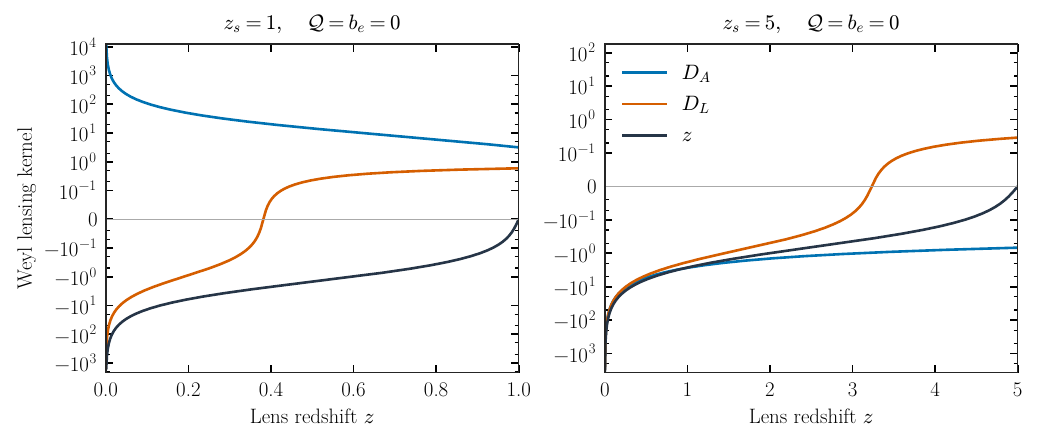}
 \caption{Weyl-potential lensing kernels for area distance in eq.~\eqref{eq:lensing-integral} and for luminosity distance and redshift from the brackets given below that equation, for $z_s=1$ and $5$ with $\Q=b_e=0$. The vertical axes use an asinh scale. The radial displacement gives the distance kernels a non-zero endpoint term.}
 \label{fig:lensing}
\end{figure}

Lensing changes the angular area assigned to a source in all three spaces. It also changes the measured distances, so a distance-selected cell moves radially as well as changing its transverse area. The radial variation of this displacement introduces a derivative of the convergence:
\begin{align}
 \left.\Delta_A\right|_\kappa&=(\beta_A-2)\kappa
                        +\frac{r}{1-r\HH}\frac{\dd\kappa}{\dd r},
 \label{eq:lensingA}\\
 \left.\Delta_L\right|_\kappa&=(\beta_L-2)\kappa
                        +\frac{r}{1+r\HH}\frac{\dd\kappa}{\dd r},
 \label{eq:lensingL}\\
 \left.\Delta_z\right|_\kappa&=2(\Q-1)\kappa.
 \label{eq:lensingz}
\end{align}
Written as a single line-of-sight integral, the area-distance term is
\begin{equation}
 \left.\Delta_A\right|_\kappa
 =\frac1{r_s}\int_0^{r_s}\dd r\,
 \left[(\beta_A-2)\frac{r_s-r}{r}+\frac1{1-r_s\HH_s}\right]
 \nabla^2_\Omega\PhiW.
 \label{eq:lensing-integral}
\end{equation}
The luminosity-distance bracket has $\beta_L$ and $1+r_s\HH_s$ in place of $\beta_A$ and $1-r_s\HH_s$; the redshift bracket is $2(\Q-1)(r_s-r)/r$. The extra term in each distance kernel comes from differentiating the radial displacement \cite{Namikawa2021,Fonseca2023}. Figure~\ref{fig:lensing} shows the resulting differences for sources below and above the area-distance maximum.

\begin{figure}[p]
 \centering\includegraphics[width=\textwidth]{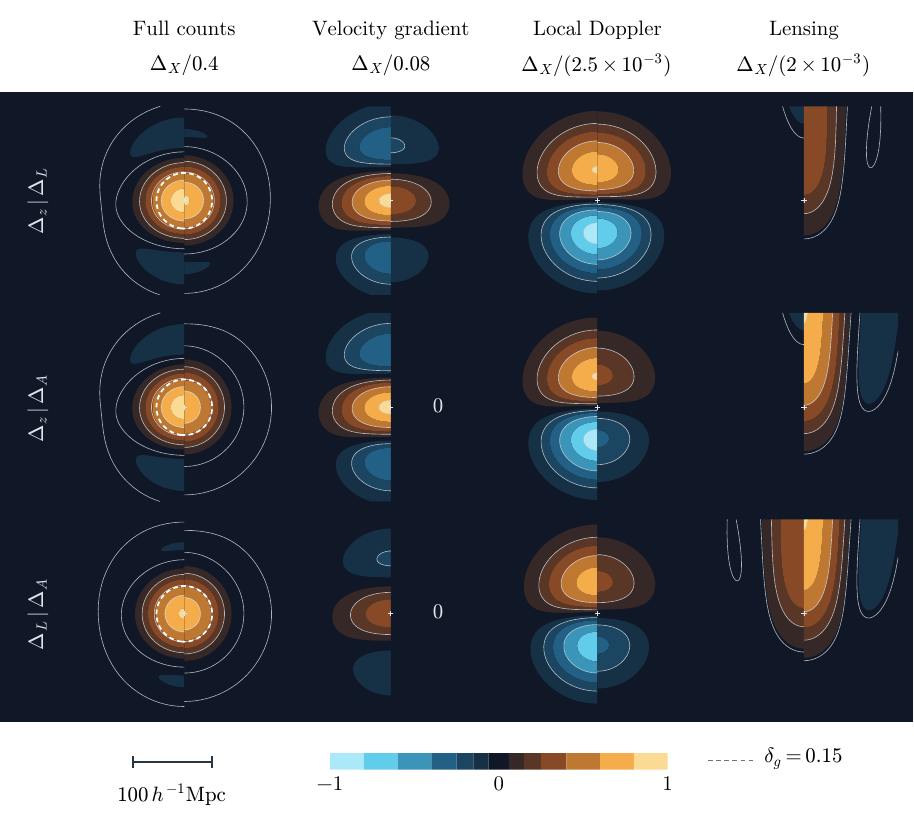}
 \caption{Full linear counts and separate contributions through the same spherical overdensity. Each map compares two distance spaces, with the left and right halves identified by the row label: redshift and luminosity distance, redshift and area distance, then luminosity and area distance. The observer lies below the panels; $r_\parallel$ and $r_\perp$ are background comoving Cartesian coordinates parallel and transverse to the central sightline. The structure is centred at $r_\parallel=r_c=r(z_c)$, $r_\perp=0$, with $z_c=0.3$. At this redshift the intrinsic profile is $\delta_g(d)=0.3[1-d^2/(3R^2)]\exp[-d^2/(2R^2)]$, where $d$ is comoving distance from the centre and $R=40\,h^{-1}{\rm Mpc}$. The density, velocity and potentials follow the same growing solution, with $\Phi=\Psi$, and are evaluated on the background lightcone. We use the background of appendix~\ref{app:numerical-spectra}, with $b=1$, $b_e=1$ and $\Q=0.5$. The same spatial scale applies to every map in both directions. Colours show the quantities below the column headings, using one stepped colour key; the physical amplitudes and contour levels are common to all examples within each column. The two halves are drawn from the same reflection-symmetric structure. In the first column, the dashed contour marks intrinsic $\delta_g=0.15$. The velocity gradient flattens the redshift and luminosity-distance counts and vanishes for area distance. Local Doppler terms have opposite signs on the near and far sides; their coefficients include the Euler and growing-mode substitutions in eqs.~\eqref{eq:velocityA}--\eqref{eq:velocityz}. Lensing includes $\kappa$ and its radial derivative where present. Its distance-space contribution extends behind the structure, continuing beyond the upper edge of the panels. Full counts include all terms at their calculated amplitudes.}
 \label{fig:overdensity-counts}
\end{figure}

Close behind a foreground structure, the convergence is suppressed by the small lens--source separation. Its radial derivative removes this factor, as the second term in eq.~\eqref{eq:lensing-integral} shows. Lensing changes the inferred spacing of distance-selected shells, so the radial volume term can exceed the usual convergence contribution. Below the area-distance maximum, its coefficient $r_s/(1-r_s\HH_s)$ is larger than the luminosity-distance coefficient $r_s/(1+r_s\HH_s)$. The count perturbation persists behind the structure as the local density and velocity decay. See figure~\ref{fig:overdensity-counts} for a density-map comparison \cite{PaulClarkson2026} through a compensated spherical overdensity.

\FloatBarrier

\subsection{Removing the common intrinsic density}
\label{sec:identity}

To compare the count fields, we begin with the same physical source events on one observer's past lightcone. Their measured radial coordinates differ, so the count density per radial interval changes. For $X,Y\in\{D_A,D_L,z\}$, local count conservation gives
\begin{equation*}
 \bar N_X(X_s)[1+\Delta_X(\bn,X_s)]\,|\dd X_s|
 =\bar N_Y(Y_s)[1+\Delta_Y(\bn,Y_s)]\,|\dd Y_s|.
\end{equation*}
\Needspace{7\baselineskip}
\noindent Here $X_s,Y_s$ are the measured coordinates of the same event. With a common background label $z$, write $X_s=\bar X(z)+\delta X$ and $Y_s=\bar Y(z)+\delta Y$. The means obey $\bar N_X(\bar X(z))|\dd\bar X/\dd z|=\bar N_z(z)$. Expanding the means and radial Jacobians gives
\begin{equation}
 \Delta_X-\Delta_Y
 =-\frac1{\bar N_z}\frac{\dd}{\dd z}
 \left[\bar N_z\left(
 \frac{\delta X}{\dd\bar X/\dd z}
 -\frac{\delta Y}{\dd\bar Y/\dd z}\right)\right].
 \label{eq:count-remapping}
\end{equation}
Each ratio measures a displacement of the background redshift label. The derivative acting on that displacement describes the change in radial interval; the derivative acting on $\bar N_z$ describes the change in the selected mean. Intrinsic density, lightcone volume and selection at the common source event are shared by the two counts and cancel in their difference.
We evaluate the first-order fields at the background coordinates, since shifting their arguments changes them only at second order. The corresponding shifts of the background means are first order and have already been included in eq.~\eqref{eq:count-remapping}.

Distance duality relates the shifts themselves,
\begin{equation}
 \frac{\delta D_L}{\dd\bar D_L/\dd z}
 -\frac{\delta D_A}{\dd\bar D_A/\dd z}
 =\frac{2r\HH}{1+r\HH}
 \left(\delta z-\frac{\delta D_A}{\dd\bar D_A/\dd z}\right).
 \label{eq:shift-identity}
\end{equation}
The bracket is the redshift shift at fixed area distance. Equation~\eqref{eq:shift-identity} therefore relates the displacement between $D_L$ and $D_A$ to the displacement between $z$ and $D_A$, with relative weight $2r\HH/(1+r\HH)$.

We now apply count conservation~\eqref{eq:count-remapping} to these two pairs. Subtracting the weighted $z$--$D_A$ count difference from the $D_L$--$D_A$ difference cancels the derivatives of both the displacement and the selected mean. The derivative of the weight remains, giving
\begin{align}
 \Delta_L-\frac{2r\HH}{1+r\HH}\Delta_z
          -\frac{1-r\HH}{1+r\HH}\Delta_A
 &=-\frac{2(1-r\HH'/\HH)}{(1+r\HH)^2}
     \left.\frac{\delta z}{1+z}\right|_{D_A}
 \nonumber\\
 &=-\frac{2(1-r\HH'/\HH)}{(1+r\HH)^2}
 \biggl\{v_r-\frac{\bv_o\cdot\bn}{1-r\HH}-\Psi
          +\frac{r\HH}{1-r\HH}\Phi \nonumber\\
 &\qquad-2\int_0^r\dd\tilde r\,\PhiW'
          -\frac{2\HH}{1-r\HH}\int_0^r\dd\tilde r\,\PhiW \nonumber\\
 &\qquad+\frac{\HH}{1-r\HH}\int_0^r\dd\tilde r\,
                \frac{r-\tilde r}{\tilde r}\nabla^2_\Omega\PhiW\biggr\}.
 \label{eq:count-identity}
\end{align}
The common intrinsic density cancels between the count fields, while the cancellation of the selected-mean derivative removes the explicit dependence on $b_e$ and $\Q$. The identity holds at linear order for the same population and selection in the flat FLRW background. It uses distance duality and count conservation, with no assumption about source motion.
The redshift perturbation here is evaluated at fixed area distance, as in eq.~\eqref{eq:fixedDA-redshift}. Peculiar velocities perturb the source redshift, while lensing and potentials shift the event selected at fixed area distance. Section~\ref{sec:spectra} compares these contributions.

\Needspace{10\baselineskip}
\section{The local kinematic dipole}
\label{sec:dipole}

A moving observer assigns different directions, distances and fluxes to the same sources. This moves sources across the boundaries of fixed observed radial bins and across the flux threshold, producing a count dipole. The related dipole in luminosity distance at fixed observed redshift is studied in ref.~\cite{BonvinDistanceDipole2006}. Following ref.~\cite{MaartensDipole2018}, we use these transformations to derive the kinematic dipole, which we write as
\begin{equation}
 \Delta_X^{\rm kin}=\Dip_X\,\bv_o\cdot\bn.
 \label{eq:dipole-definition}
\end{equation}
We compare observers at the same reception event, holding the source worldlines and lightcone fixed. Only the measured coordinates and fluxes change; each source emission event is the same. The observer scalar terms in appendix~\ref{app:observer} are unchanged. We retain terms linear in the boost and omit its products with cosmological perturbations.

\subsection{Aberration, radial shifts and selection}

To first order in the boost, the observed intervals and coordinates transform as
\begin{align}
 \dd{\widetilde\Omega_o} &=(1-2\bv_o\cdot\bn)\dd{\Omega_o},
 &1+\widetilde z&=(1+z)(1-\bv_o\cdot\bn),\nonumber\\
 \widetilde D_A&=D_A(1+\bv_o\cdot\bn),
 &\widetilde D_L&=D_L(1-\bv_o\cdot\bn).
 \label{eq:boost}
\end{align}
Forward aberration reduces the observed solid angle of a fixed source area, so the inferred $D_A$ increases. The simultaneous decrease in redshift reverses the distance shift in $D_L=(1+z)^2D_A$. The bolometric flux consequently increases by the fractional amount $2\bv_o\cdot\bn$.

Conservation of a fixed set of sources in the transformed cell gives
\begin{align}
 \widetilde N_A(\widetilde D_A)&=(1+\bv_o\cdot\bn)N_A(D_A),\nonumber\\
 \widetilde N_L(\widetilde D_L)&=(1+3\bv_o\cdot\bn)N_L(D_L),\nonumber\\
 \widetilde N_z(\widetilde z)&=(1+3\bv_o\cdot\bn)N_z(z).
 \label{eq:raw-dipoles}
\end{align}
Aberration contributes $+2$ in each space. The radial interval contributes $-1$ for $D_A$ and $+1$ for $D_L$ and $z$.

To form a map at a fixed boosted coordinate, we must also account for the change in the selected shell. For area distance, the unboosted shell lies at the smaller distance $D_A=\widetilde D_A(1-\bv_o\cdot\bn)$. A fixed observed flux cut also admits intrinsically fainter sources in the forward direction, giving
\begin{align}
 \widetilde N_A(\widetilde D_A;F_{\rm cut})
 ={}&(1+\bv_o\cdot\bn)\,
 \bar N_A\bigl(\widetilde D_A[1-\bv_o\cdot\bn];
                    F_{\rm cut}[1-2\bv_o\cdot\bn]\bigr)
 +O(v_o^2).
 \label{eq:fixed-cell-boost}
\end{align}
Expanding about $\bar N_A(\widetilde D_A;F_{\rm cut})$ gives the fractional count perturbation $[1-\dd\ln\bar N_A/\dd\ln D_A+2\Q]\bv_o\cdot\bn$: the slope of the mean counts describes the shell displacement, and $\Q$ describes the additional sources admitted by the flux shift. The inverse shifts of $D_L$ and $z$ have the opposite sign, so their mean-slope terms enter positively. 

Using the background means
\begin{equation}
 \bar N_z=\frac{\bar n_{\rm sel}r^2}{H},\qquad
 \bar N_L=\frac{a\bar n_{\rm sel}r^2}{1+r\HH},\qquad H=\frac\HH a,
 \label{eq:other-means}
\end{equation}
we obtain the dipole amplitudes
\begin{align}
 \Dip_A&=1-\frac{\dd\ln\bar N_A}{\dd\ln D_A}+2\Q
          =2-\beta_A+4\Q,\nonumber\\
 \Dip_L&=3+\frac{\dd\ln\bar N_L}{\dd\ln D_L}+2\Q
          =2+\beta_L,\nonumber\\
 \Dip_z&=3+(1+z)\frac{\dd\ln\bar N_z}{\dd z}+2\Q
          =2+\frac{\HH'}{\HH^2}+\frac{2-2\Q}{r\HH}-b_e.
 \label{eq:dipoles}
\end{align}
The slopes include the selected evolution in eq.~\eqref{eq:selected-evolution}. Equation~\eqref{eq:countsA} gives the same area-distance coefficient: $2$ from aberration, $-\beta_A$ from distance and $4\Q$ from redshift. 
The redshift result has the form of ref.~\cite{MaartensDipole2018}, with $5s=2\Q$ and $\HH'/\HH^2=1+\dot H/H^2$, provided $b_e$ is defined at fixed intrinsic luminosity threshold as in ref.~\cite{MaartensBias2021}. A derivative along a fixed apparent-flux limit includes the additional threshold evolution in eq.~\eqref{eq:selected-evolution} and cannot be substituted for $b_e$ without converting the selection term.

\begin{table}[t]
 \centering
 \begin{tabular}{lccc}
 \toprule
 Space & Geometry & Evolution & Flux selection\\
 \midrule
 $D_A$ & $\displaystyle\frac{r^2(2\HH^2+\HH')-2r\HH-1}{(1-r\HH)^2}$
       & $\displaystyle\frac{r\HH b_e}{1-r\HH}$
       & $\displaystyle\frac{4\Q}{1-r\HH}$\\[9pt]
 $D_L$ & $\displaystyle\frac{r^2(2\HH^2+\HH')+6r\HH+5}{(1+r\HH)^2}$
       & $\displaystyle-\frac{r\HH b_e}{1+r\HH}$ & $0$\\[9pt]
 $z$ & $\displaystyle 2+\frac{\HH'}{\HH^2}+\frac2{r\HH}$ & $-b_e$
       & $\displaystyle-\frac{2\Q}{r\HH}$\\[3pt]
 \bottomrule
 \end{tabular}
 \caption{The three contributions to the kinematic dipole coefficients. Geometry includes aberration, the radial Jacobian and the change in background volume. Evolution is at fixed intrinsic luminosity threshold.}
 \label{tab:dipoles}
\end{table}

Table~\ref{tab:dipoles} separates these contributions. Source evolution follows from the emission-time shifts at fixed observed coordinate,
\begin{equation}
 \left.\HH\delta\eta_s\right|_{D_A}
 =\frac{r\HH}{1-r\HH}\bv_o\cdot\bn,\qquad
 \left.\HH\delta\eta_s\right|_{D_L}
 =-\frac{r\HH}{1+r\HH}\bv_o\cdot\bn,\qquad
 \left.\HH\delta\eta_s\right|_z=-\bv_o\cdot\bn.
 \label{eq:source-time-boost}
\end{equation}
Multiplying by $b_e$ gives the evolution column. At fixed $D_L$, the threshold $4\pi F_{\rm cut}D_L^2$ is fixed. At fixed $D_A$, its factor $(1+z)^4$ changes, and at fixed $z$ its factor $D_A^2$ changes. The corresponding shifts are
\begin{equation}
 \left.\delta\ln L_{\rm cut}\right|_{D_A}
 =-\frac{4\bv_o\cdot\bn}{1-r\HH},\qquad
 \left.\delta\ln L_{\rm cut}\right|_{D_L}=0,\qquad
 \left.\delta\ln L_{\rm cut}\right|_z
 =\frac{2\bv_o\cdot\bn}{r\HH}.
 \label{eq:threshold-boost}
\end{equation}
Multiplying by $-\Q$ gives the selection column of the table.

\subsection{Why the dipoles can have opposite signs}

The dipoles can reverse because the same boost changes which radial shell is selected. A forward boost raises the measured area distance but lowers luminosity distance and redshift. On the increasing branch of $\bar D_A$, a fixed $D_A$ bin therefore selects a nearer shell, while fixed $D_L$ and $z$ bins select farther shells. For finite biases at small $r\HH$, the coefficients become
\begin{equation}
 \Dip_A\longrightarrow-1+4\Q,\qquad
 \Dip_L\longrightarrow5,\qquad
 \Dip_z=\frac{2-2\Q}{r\HH}+O(1).
 \label{eq:dipole-limits}
\end{equation}
The geometrical part is easiest to see nearby, where the background volume per unit distance grows as the square of the distance. At fixed $D_A$, the nearer shell selected by a forward boost contains less volume; this loss exceeds the $+1$ Jacobian contribution in eq.~\eqref{eq:raw-dipoles}. At fixed $D_L$, the selected shell is farther away, so the extra volume adds to the $+3$ Jacobian contribution. Flux selection then changes these geometrical coefficients as in eq.~\eqref{eq:dipole-limits}. 

\begin{figure}[t]
 \centering\includegraphics[width=\textwidth]{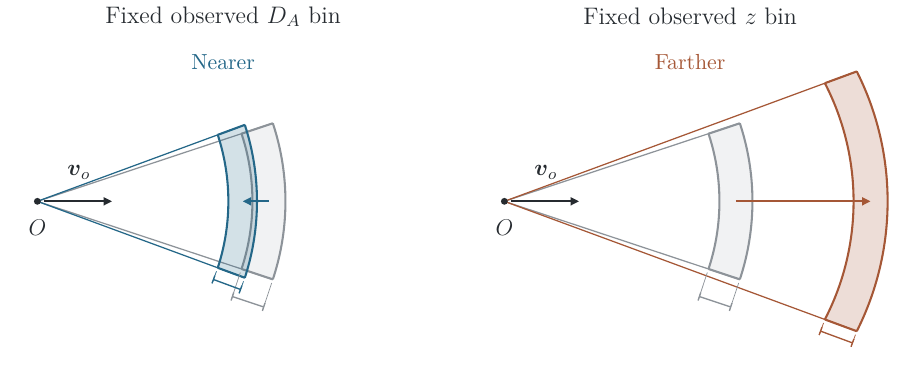}
 \caption{Source volumes selected by fixed observed distance and angular bins before (grey) and after (colour) an observer boost to the right, drawn in unboosted spatial coordinates. Aberration reduces the observed solid angle of a fixed bundle of rays: $\dd\widetilde\Omega_o=(1-2\bv_o\cdot\bn)\dd\Omega_o$. A fixed boosted aperture therefore covers the larger unboosted angle $\dd\Omega_o=(1+2\bv_o\cdot\bn)\dd\widetilde\Omega_o$, giving the wider coloured cones. On the increasing branch of $\bar D_A$, the boost selects a nearer shell at fixed $D_A$ and a farther shell at fixed $z$. Nearby, the smaller $D_A$ radius cancels the larger selected solid angle in $r^2\dd\Omega_o$; the reduced radial depth then gives the volume-limited coefficient $\Dip_A=-1$. Brackets mark the radial depths. The displacements are exaggerated.}
 \label{fig:observer-shell-selection}
\end{figure}

In redshift space the fractional shift scales as $v_o/z$, enhancing the nearby count dipole unless $\Q=1$ \cite{MaartensDipole2018}. The same inverse-redshift scaling occurs in the observer-motion contribution to the expansion-rate dipole \cite{Kalbouneh2024,Kalbouneh2026}. The linear expansion requires this shift to remain small relative to the scale over which the selected mean varies. See figure~\ref{fig:observer-shell-selection} for the shell displacements and aberration.

\begin{figure}[t]
 \centering\includegraphics[width=\textwidth]{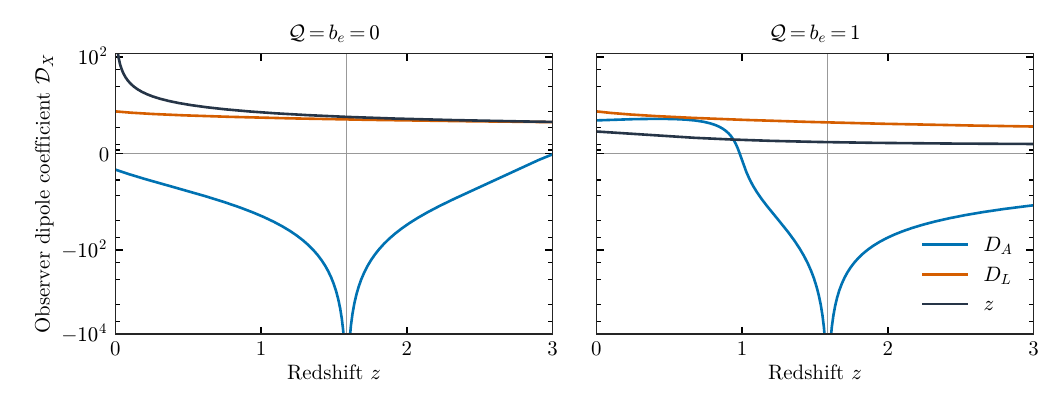}
 \caption{Kinematic dipole coefficients in eq.~\eqref{eq:dipoles}, with the bias choices and colours of figure~\ref{fig:velocity}. A negative coefficient reverses the dipole relative to $\bv_o$. The vertical axes use an asinh scale. The grey line marks the area-distance maximum, where the pointwise expansion fails.}
 \label{fig:dipole}
\end{figure}

The area-distance dipole can be much larger than the luminosity-distance and redshift dipoles. At $z=1.2$, $|\Dip_A|$ exceeds both $|\Dip_L|$ and $|\Dip_z|$ for each bias choice in figure~\ref{fig:dipole}. This enhancement comes from the small radial derivative of $\bar D_A$: a fixed boost shifts the selected shell farther along the lightcone. The effect grows towards the area-distance maximum, where the linear radial expansion eventually fails. Near the observer, redshift space can instead have the largest dipole, as eq.~\eqref{eq:dipole-limits} shows.
Measuring the larger area-distance signal requires distance information for many sources. Distance scatter limits the radial resolution, especially near the maximum. Section~\ref{sec:size-magnification} discusses the statistical information available from galaxy sizes.

\Needspace{9\baselineskip}
\subsection{Ellis--Baldwin formula and radial cuts}
\label{sec:ellis-baldwin}

Although the dipole coefficients differ between radial coordinates, integrating the counts over the same complete catalogue must give the same projected angular dipole. To see how the radial selection enters, we rewrite eq.~\eqref{eq:dipoles} for area distance as
\begin{equation}
 \bar N_A\Dip_A=(2+2\Q)\bar N_A
              -\frac{\dd}{\dd D_A}(D_A\bar N_A).
 \label{eq:angular-dipole}
\end{equation}
The derivative becomes a boundary term on integration: it counts the sources that a boost moves across the distance cuts. Away from the distance maximum, the count-weighted fluctuation in a fixed observed bin is
\begin{equation}
 \Delta_{A,[D_1,D_2]}^{\rm kin}
 =\left[2+2\langle\Q\rangle_{\rm bin}
       -\frac{[D_A\bar N_A]_{D_1}^{D_2}}
              {\displaystyle\int_{D_1}^{D_2}\bar N_A\,\dd D_A}\right]
       \bv_o\cdot\bn,
 \label{eq:binned-area-dipole}
\end{equation}
where $\langle\Q\rangle_{\rm bin}$ is weighted by $\bar N_A$. At low redshift, with no flux selection and $\bar N_A\propto D_A^2$, the boundary term contributes $-3$, leaving the coefficient $2-3=-1$. Observer motion moves sources across luminosity-distance and redshift cuts in the opposite radial direction, as given by eq.~\eqref{eq:boost}.

When the angular catalogue includes the full admitted radial range, changing the radial coordinate moves no source into or out of that range. Count conservation leaves $2+2\langle\Q\rangle$ in all three spaces, the bolometric Ellis--Baldwin coefficient \cite{EllisBaldwin1984,vonHausegger2024}. Source evolution is already included in the observed count distribution and flux-threshold slope. 
In area distance, the complete catalogue must include every admitted branch. Since $D_A\bar N_A$ diverges at the distance maximum, eq.~\eqref{eq:angular-dipole} cannot be integrated by parts through that point; the counts must instead be integrated in a regular source coordinate as in appendix~\ref{app:branches}.
Boundary terms and sign reversals for redshift cuts and general windows are derived in ref.~\cite{vonHauseggerDalang2025}. A photometric flux cut also requires the spectral and bandpass response in place of the bolometric boost \cite{Bonnefous2026}.

For a prescribed observer velocity, the kinematic contribution to the $\ell=1$ multipole of the angular cross-spectrum is
\begin{equation}
 C_1^{XY,\rm kin}(z,z')
 =\frac{4\pi}{9}v_o^2\Dip_X(z)\Dip_Y(z').
 \label{eq:cross-dipole}
\end{equation}
Opposite dipole directions give negative cross-power. If the observer velocity is treated as stochastic, its correlations with the source fields also contribute.

\subsection{A combination independent of evolution and flux selection}

The dependence of each dipole on evolution and flux selection can be removed by combining the three count fields. Applying eq.~\eqref{eq:count-identity} to the observer boost gives
\begin{align}
 \Delta_L^{\rm kin}
 -\frac{2r\HH}{1+r\HH}\Delta_z^{\rm kin}
 -\frac{1-r\HH}{1+r\HH}\Delta_A^{\rm kin}
 =\frac{2(1-r\HH'/\HH)}{(1-r\HH)(1+r\HH)^2}\bv_o\cdot\bn.
 \label{eq:dipole-combination}
\end{align}
The observer-velocity coefficient depends only on the background geometry. The same count combination also contains the cosmological dipole of the redshift perturbation in eq.~\eqref{eq:count-identity}, including source velocities and gravitational contributions. These contribute to the same angular multipole and must be included when using the combination to estimate the observer velocity.

\section{Angular power spectra}
\label{sec:spectra}

The absence of the Kaiser term and the different lensing contributions change both the power in each count field and its correlation with the others. We compare these effects using full-sky spectra at the same background source positions. The notation $\Delta_X(\bn,z)$ denotes the field at $\bar X(z)$, with $z$ serving as the background label as in section~\ref{sec:identity}. Its spherical-harmonic coefficients satisfy
\begin{align}
 \langle a_{\ell m}^X(z)a_{\ell'm'}^{Y*}(z')\rangle
 &=\delta_{\ell\ell'}\delta_{mm'}C_\ell^{XY}(z,z'),\nonumber\\
 C_\ell^{XY}(z,z')
 &=\frac2\pi\int_0^\infty\dd k\,k^2P_0(k)
       T_\ell^X(k,z)T_\ell^Y(k,z').
 \label{eq:angular-spectra}
\end{align}
Here $P_0$ is the present comoving-gauge matter power spectrum and $T_\ell^X$ is the count transfer function relative to that matter contrast.

Density and radial velocity project onto $bD_+j_\ell$ and $(\HH fD_+/k)j_\ell'$, respectively. Here the Bessel functions are evaluated at $kr$, and their primes denote derivatives of the argument. The Kaiser terms project onto $-fD_+j_\ell''$ in redshift space and $-2r\HH fD_+j_\ell''/(1+r\HH)$ in luminosity distance space; the corresponding term is absent in area distance. Projecting eqs.~\eqref{eq:distance-perturbation}--\eqref{eq:countsA} and using count conservation in eq.~\eqref{eq:count-remapping} supplies the remaining local and integrated terms.

For the numerical examples we use a flat matter-plus-$\Lambda$ model and an illustrative population with $b=1$, $b_e=1$ and $\Q=0.5$. Appendix~\ref{app:numerical-spectra} specifies the matter spectrum, Fourier cutoff and plotted ranges. The comparisons below use the same source positions and Fourier selection in all three radial coordinates.

\subsection{Full spectra and differences between radial coordinates}

\begin{figure}[t]
 \centering\includegraphics[width=\textwidth]{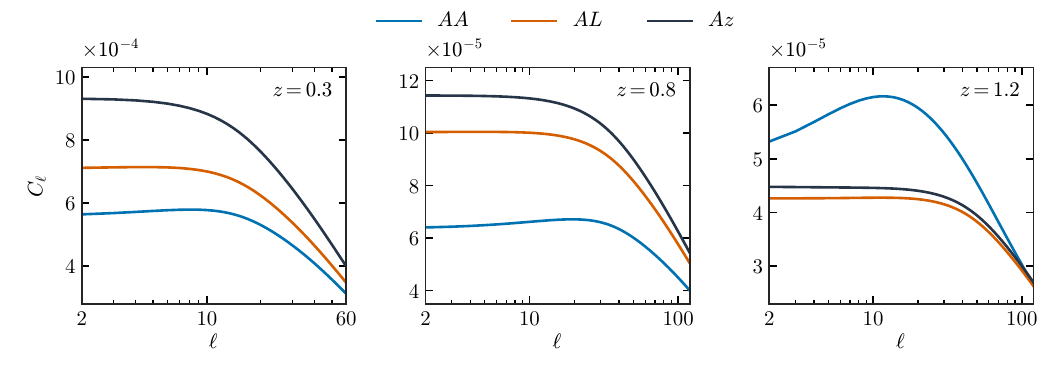}
 \caption{Area-distance auto- and cross-spectra at the same background source position, plotted as $C_\ell$. The illustrative population has $b=1$, $b_e=1$ and $\Q=0.5$, with no radial averaging. The common Fourier selection and the plotted ranges are described in appendix~\ref{app:numerical-spectra}.}
 \label{fig:spectra}
\end{figure}

The three fields share the same intrinsic density, which keeps them strongly correlated at low redshift. Radial displacements and light propagation add different contributions in each space, changing both their auto- and cross-spectra in figure~\ref{fig:spectra}. For the chosen population, the correlation coefficient $C_2^{Az}/\sqrt{C_2^{AA}C_2^{zz}}$ falls from $1.00$ at $z=0.3$ to $0.66$ at $z=1.2$.

\begin{figure}[tp]
 \centering\includegraphics[width=\textwidth]{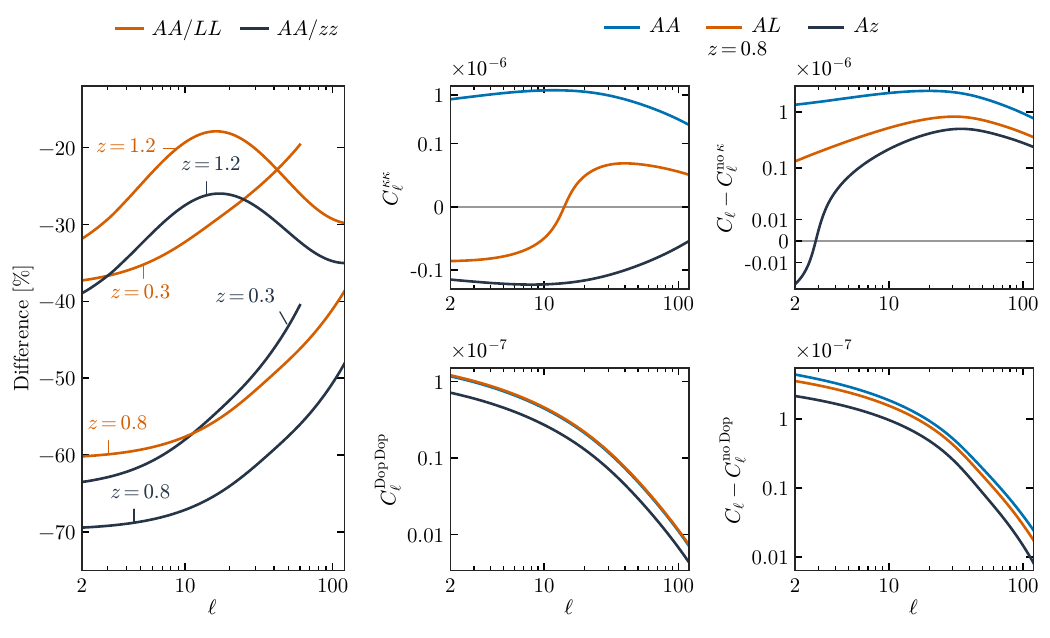}
 \caption{Effect of the radial coordinate, lensing and local Doppler terms, for the population in figure~\ref{fig:spectra}. Left: relative area-distance auto-power, $100(C_\ell^{AA}/C_\ell^{XX}-1)$ for $X=L,z$, at the three redshifts labelled on the curves. The four panels on the right show $z=0.8$: the middle column gives the spectra of the isolated lensing terms (top) and Doppler terms (bottom), while the right column gives the change in the full spectra when those terms are removed. These changes include cross-correlations with the remaining count terms. Lensing includes $\kappa$ and its radial derivative. Doppler includes the terms proportional to $v_r$ in eqs.~\eqref{eq:velocityA}--\eqref{eq:velocityz} after the Euler and growing-mode substitutions; the Kaiser gradients remain in the rest of the field. The spectral panels show $AA$, $AL$ and $Az$, with the colours of figure~\ref{fig:spectra}. Their vertical axes are logarithmic where all curves are positive and asinh otherwise. The $z=0.3$ curves stop at $\ell=60$ and the other curves at $\ell=120$; a common Fourier taper removes modes above $k_{\max}=0.16\,{\rm Mpc}^{-1}$ to limit nonlinear contributions. Figure~\ref{fig:coordinate-lensing-appendix} shows the component breakdowns at $z=0.3$ and $1.2$, with cutoff details in appendix~\ref{app:numerical-spectra}.}
 \label{fig:space-differences}
 \label{fig:kappa-breakdown}
 \label{fig:doppler-breakdown}
\end{figure}

The area-distance auto-power is lower than in either luminosity distance or redshift space, with the largest deficit at $z=0.8$ in these examples. At $z=1.2$, the larger lensing contribution partly offsets the missing velocity-gradient power, reducing the deficit relative to $z=0.8$ throughout the plotted range. Figure~\ref{fig:space-differences} compares the three redshifts and shows the lensing and Doppler contributions at $z=0.8$; figure~\ref{fig:coordinate-lensing-appendix} gives the corresponding component breakdowns at $z=0.3$ and $1.2$.

\subsection{Velocity and optical contributions}

In luminosity distance and redshift space, the velocity gradient adds power through both its auto-correlation and its correlation with density. These terms are absent from $C_\ell^{AA}$. At larger distances, lensing becomes comparable to density in the area-distance spectrum, while the individual potential contributions remain smaller. The component auto-powers and their cross-correlations are shown in appendix~\ref{app:numerical-spectra}.

The Doppler panels in figure~\ref{fig:space-differences} isolate the local velocity terms in eqs.~\eqref{eq:velocityA}--\eqref{eq:velocityz}, including the potential gradient absorbed by the Euler and growing-mode substitutions. Their auto-power can be small even when their correlations with the rest of the count field change the full spectrum appreciably. At $z=1.2$, removing these terms changes $C_2^{AA}$ by $2.5\%$ (figure~\ref{fig:coordinate-lensing-appendix}).

Lensing has a much larger effect at this distance. Removing convergence lowers $C_2^{AA}$ by $53\%$ and $C_{120}^{AA}$ by $31\%$, and raises the correlation coefficient with redshift space from $0.66$ to $0.98$ at $\ell=2$. Lensing therefore both increases the area-distance power and weakens its correlation with redshift counts.
Population selection also matters: at $z=1.2$, the chosen $b_e=1$, $\Q=0.5$ lower $C_2^{AA}$ by $37\%$ relative to $b_e=\Q=0$. 

The combination in eq.~\eqref{eq:count-identity} removes the intrinsic density and the explicit dependence on $b_e$ and $\Q$. It is proportional to the redshift perturbation at fixed area distance, which is dominated by source velocities at $z=0.3$, with optical terms taking over by $z=0.8$ in these examples. The separate spectra are given in appendix~\ref{app:fixed-distance-spectrum}; section~\ref{sec:size-magnification} relates this signal to galaxy-size magnification.

\Needspace{9\baselineskip}
\section{Distance information and geometrical tests}
\label{sec:applications}

To measure clustering in area distance space, we need distance information beyond the observed redshift. Joint measurements of $D_L$ and $z$ give $D_A=D_L/(1+z)^2$, while galaxy-size correlations provide a statistical measure of distance perturbations. Assigning $\bar D_A(z_{\rm obs})$ using the observed redshift alone simply relabels the redshift-space positions, including their velocity distortions. 

Distance errors set the radial resolution. For small errors, the corresponding radial uncertainties are
\begin{equation}
 \sigma_{r,A}\simeq\frac{r\,\sigma_{\ln D_A}}{|1-r\HH|},
 \qquad
 \sigma_{r,L}\simeq\frac{r\,\sigma_{\ln D_L}}{1+r\HH}.
 \label{eq:distance-error-propagation}
\end{equation}
When redshift errors are negligible, $\sigma_{\ln D_A}=\sigma_{\ln D_L}$ for $D_A=D_L/(1+z)^2$. Area-distance bins then have poorer radial resolution, especially near the distance maximum. Broad distance errors require averaging the counts over the distance-error distribution.

\subsection{Galaxy sizes and Doppler magnification}
\label{sec:size-magnification}

For a galaxy with area-equivalent physical radius $R$, the angular radius obeys $\theta=R/D_A$. Distance perturbations therefore change its apparent size. Galaxies have a broad distribution of physical sizes, so we infer magnification statistically using a calibrated size estimator and its correlations \cite{Huff2014,Alsing2015,SchmidtMagnification2012}. At fixed observed redshift, the distance perturbation in eq.~\eqref{eq:distance-at-observed-redshift} gives the total convergence,
\begin{align}
 \kappa_{\rm tot}
 &\equiv-\frac{\left.\delta D_A\right|_z}{\bar D_A}
 =\kappa_{\rm Dop}+\kappa+\Phi
       -\frac2r\int_0^r\dd\tilde r\,\PhiW\nonumber\\
 &\quad-\frac{1-r\HH}{r\HH}
       \left[\Psi+2\int_0^r\dd\tilde r\,\PhiW'\right],
 \label{eq:magnification-redshift}
\end{align}
where $\kappa$ is the gravitational convergence of eq.~\eqref{eq:kappa}. 
Equation~\eqref{eq:fixedDA-redshift} gives $\delta z|_{D_A}/(1+z)=r\HH\kappa_{\rm tot}/(1-r\HH)$, so the bias-independent count combination in eq.~\eqref{eq:count-identity} is proportional to this magnification field. The distance--redshift relation, and hence magnification, has been derived at first \cite{Bonvin2006,HuiGreene2006} and second order \cite{Umeh2014I,Umeh2014II}.

At fixed observed redshift $z_{\rm obs}$, the Doppler contribution is
\begin{equation}
 \kappa_{\rm Dop}(\bn,z_{\rm obs})
 =\left(\frac1{r\HH}-1\right)v_r
  -\frac{\bv_o\cdot\bn}{r\HH},
 \qquad r=r(z_{\rm obs}).
 \label{eq:doppler-convergence}
\end{equation}
The background factors and source fields are evaluated at the reference time for $z_{\rm obs}$. Although source motion leaves the beam area at a fixed event unchanged, it moves the event selected by a fixed observed redshift. The resulting change in distance alters the apparent size and produces Doppler magnification \cite{Bonvin2008,Bolejko2013,Bacon2014,BonvinDipole2017,Andrianomena2019}. Infall makes galaxies appear larger on the near side of an overdensity and smaller on its far side, producing a dipole in the density--size correlation \cite{Bacon2014,BonvinDipole2017}. Outflow around voids reverses the pattern \cite{Bolejko2013}. This correlation provides a statistical measure of peculiar velocities and a route to tests of gravity \cite{Andrianomena2019}.

\Needspace{11\baselineskip}
The magnification field also determines the radial displacement between redshift and area-distance bins. Substituting the fixed-redshift distance perturbation, $\delta D_A|_z=-\bar D_A\kappa_{\rm tot}$, into count conservation gives
\begin{align}
 \Delta_A\bigl(\bn,\bar D_A(z)\bigr)-\Delta_z(\bn,z)
 &=\frac1{\bar N_z}\frac{\dd}{\dd z}
 \left[
 \bar N_z\,\frac{\bar D_A}{\dd\bar D_A/\dd z}\,
 \kappa_{\rm tot}(\bn,z)
 \right]\nonumber\\
 &=\frac r{1-r\HH}\frac{\dd\kappa_{\rm tot}}{\dd r}
 +\left[
 \frac{3-2r\HH-r^2\HH'}{(1-r\HH)^2}
 -\frac{r\HH b_e+2\Q(1+r\HH)}{1-r\HH}
 \right]\kappa_{\rm tot}.
 \label{eq:counts-from-magnification}
\end{align}
The coefficient of $\kappa_{\rm tot}$ is $\beta_A-2\Q$, and $\dd/\dd r$ follows the background lightcone. Expanding in velocities and potentials, with $\Phi$ and $\Psi$ independent and before applying the source Euler equation, gives
\begin{align}
 \Delta_A-\Delta_z={}&
 \frac1\HH(\partial_r v_r-v_r'-\partial_r\Psi)
 +\frac r{1-r\HH}\partial_r\Phi
 -\frac{\Phi'}{\HH(1-r\HH)}\nonumber\\
 &+\left[\frac{\HH'}{\HH^2}+\frac{2-2\Q}{r\HH}-2\Q-b_e\right]
 \left[v_r-\bv_o\cdot\bn-\Psi
                  -2\int_0^r\dd\tilde r\,\PhiW'\right]\nonumber\\
 &+(\beta_A-2\Q)
    \left[\Phi-\bv_o\cdot\bn-\frac2r\int_0^r\dd\tilde r\,\PhiW\right]
    -\frac{\Phi+\Psi}{1-r\HH}\nonumber\\
 &+\frac2{r(1-r\HH)}\int_0^r\dd\tilde r\,\PhiW\nonumber\\
 &+\frac1r\int_0^r\dd\tilde r\,
     \left[(\beta_A-2\Q)\frac{r-\tilde r}{\tilde r}
                        +\frac1{1-r\HH}\right]\nabla^2_\Omega\PhiW.
 \label{eq:size-count-expanded}
\end{align}
For the same population and selection, the area-distance correlations can therefore be assembled from the redshift-count correlations, the magnification correlations and their cross-correlations, with the radial derivatives in eq.~\eqref{eq:counts-from-magnification} acting on them. Spectroscopic redshifts specify where these correlations are measured; size correlations supply the additional distance information without requiring an accurate ruler for each galaxy.

To infer $\kappa_{\rm tot}$ from mean log-size, the estimator must be calibrated for size and flux cuts and for physical-size evolution. Intrinsic scatter and measurement noise must also be included \cite{Alsing2015}. 
A recent LSST--DESI forecast \cite{Ye2026} combines spectroscopic density measurements with galaxy sizes from simulated LSST images. It finds that intrinsic size scatter dominates over measurement noise, but still forecasts a detectable Doppler magnification dipole in several low-redshift bins. Broad intrinsic scatter requires a distribution of estimated distances, averaged over when predicting the count correlations.

Intrinsic sizes and Fundamental Plane residuals also correlate with density \cite{Joachimi2015,Ciarlariello2015,Ciarlariello2016}. The density--size pair dipole suppresses the leading symmetric contribution \cite{BonvinDipole2017,Andrianomena2019,Ye2026}, but evolution, wide-angle and relativistic correlations remain \cite{Paul2023}. This pair-separation dipole differs from the observer's angular dipole in section~\ref{sec:dipole}.

\Needspace{10\baselineskip}
\subsection{Alcock--Paczynski and acoustic-ruler measurements}

The Alcock--Paczynski (AP) test compares radial and transverse clustering scales using statistical isotropy \cite{Alcock1979}. Peculiar velocities also generate anisotropy, so redshift-space analyses must distinguish their distortion from the geometrical one \cite{Ballinger1996}. BAO supplies a comoving ruler of length $r_d$ \cite{Seo2003}. The relevant transverse and radial distances and their AP ratio are
\begin{equation}
 D_M=(1+z)\bar D_A,\qquad D_H=\frac1H,\qquad
 F_{\rm AP}\equiv\frac{D_M}{D_H}=H D_M.
 \label{eq:standard-AP}
\end{equation}
The angular and radial BAO scales determine $D_M/r_d$ and $D_H/r_d$, while their ratio $F_{\rm AP}$ is independent of the ruler length. These ratios constrain the background geometry; the perturbation of each source distance requires the additional information discussed above.

Radial and transverse clustering can also be compared when a distance serves as the radial coordinate, as developed for luminosity distance in ref.~\cite{Zhang2018}. In our spatially flat background,
\begin{equation}
 \frac{\dd\bar D_A}{\dd r}=a(1-r\HH),
 \qquad
 \frac{\dd\bar D_L}{\dd r}=a^{-1}(1+r\HH).
 \label{eq:distance-ruler-derivatives}
\end{equation}
For a local isotropic comoving ruler, neglecting background evolution across its length,
\begin{equation}
 \theta_{\rm BAO}\simeq\frac{r_d}{r},
 \qquad
 \Delta X_{\rm BAO}\simeq
   \left|\frac{\dd\bar X}{\dd r}\right|r_d,
 \qquad X\in\{z,D_A,D_L\}.
 \label{eq:radial-ruler}
\end{equation}
Dividing the radial interval by the angular size eliminates $r_d$ and gives the ratios in table~\ref{tab:ruler-ratios}. The absolute value accounts for the reversal of the area-distance gradient beyond its maximum.

\begin{table}[!htbp]
 \centering
 \begin{tabular}{lcc}
 \toprule
 Radial coordinate & Ruler ratio & Flat FLRW value\\
 \midrule
 $z$ & $\Delta z_{\rm BAO}/\theta_{\rm BAO}$
     & $Hr=F_{\rm AP}$\\[5pt]
 $D_A$ & $\Delta D_{A,{\rm BAO}}/(\bar D_A\theta_{\rm BAO})$
       & $|1-r\HH|$\\[5pt]
 $D_L$ & $\Delta D_{L,{\rm BAO}}/(\bar D_L\theta_{\rm BAO})$
       & $1+r\HH$\\[3pt]
 \bottomrule
 \end{tabular}
 \caption{Local radial-to-transverse ruler ratios for the same background source position. The ruler length cancels. These are geometrical relations; velocity and optical perturbations enter through the full observed count fields.}
 \label{tab:ruler-ratios}
\end{table}

When area-distance counts are inferred from galaxy sizes, the assumed background relation $\bar D_A(z)$ also sets the coefficients that convert magnification into count fluctuations in eq.~\eqref{eq:counts-from-magnification}. An AP fit must vary these coefficients together with the background geometry. The joint count--size correlations must include optical terms, intrinsic size correlations and measurement errors. A constant distance calibration cancels from the normalised ruler ratios, while a varying calibration changes the radial derivative and can mimic a change in geometry.

\subsection{Other radial observables}
\label{sec:other-radial-observables}

Dispersion measure (DM) provides another radial coordinate for fast radio bursts. Its mean increases with distance, while its fluctuations depend on the intervening free-electron column. Linear DM-space counts also lack a Kaiser velocity-gradient term, but contain local and integrated electron-density contributions \cite{AlonsoDM2021}.

Consider the same source population and selection described using either DM or area distance. On a branch where both background relations can be inverted, count conservation~\eqref{eq:count-remapping} gives
\begin{equation}
 \Delta_{\rm DM}-\Delta_A
 =-\frac1{\bar N_z}\frac{\dd}{\dd z}
 \left[\bar N_z\left(
 \frac{\delta{\rm DM}}{\dd\overline{\rm DM}/\dd z}
 -\frac{\delta D_A}{\dd\bar D_A/\dd z}\right)\right].
 \label{eq:dm-area-counts}
\end{equation}
Here $\overline{\rm DM}(z)$ is the background mean, and both perturbations are evaluated at the same background emission time. Distance duality relates the optical distances and redshift in eq.~\eqref{eq:count-identity}. DM additionally depends on the intervening free-electron distribution, so the electron-density terms remain in the count difference. 
Cross-correlating DM with galaxy-size magnification could compare this electron distribution with gravitational focusing, allowing for electron bias, host-galaxy DM and intrinsic size correlations.

\FloatBarrier
\section{Conclusions}

We have derived the linear area-distance counts for a flux-selected population and compared them with luminosity distance and redshift space for the same sources. The absence of a Kaiser term follows from the geometry of an infinitesimal beam: its area at fixed source and observer events is independent of the source velocity at any order. There is consequently no source Doppler radial displacement, although velocities still enter the counts through the current and flux selection. Lensing changes both the angular and radial assignment of sources, increasing the area-distance power and weakening its correlation with redshift counts in the examples studied here.

A forward observer boost increases area distance while decreasing luminosity distance and redshift. On the increasing area-distance branch, a fixed radial bin therefore selects shells displaced in opposite directions, allowing the area-distance kinematic dipole to reverse relative to those in the other two spaces. These differences arise from the radial selection: radially integrating the same complete catalogue recovers the common Ellis--Baldwin result.

A weighted combination of the three count fields cancels the common intrinsic density, radial derivatives and explicit evolution and magnification biases. It is proportional to the redshift perturbation at fixed area distance, dominated by velocities near the observer and optical effects farther away in our numerical examples.
This combination is directly related to galaxy-size magnification, allowing its correlations to be inferred statistically without precise individual rulers. Joint luminosity-distance and redshift measurements provide another route to area-distance clustering. In both cases, distance errors limit the radial resolution, particularly near the area-distance maximum where a small distance error corresponds to a large change in radial position.

\section*{Acknowledgements}
CC is supported by STFC (UK) grant ST/X000931/1. CC also acknowledges OpenAI for providing complimentary access to ChatGPT and Codex through the ChatGPT for Academic Researchers programme. ChatGPT (OpenAI) and Claude (Anthropic) were used to assist with manuscript drafting and revision, algebraic checks, figure preparation and numerical code. The authors take full responsibility for the calculations, interpretation and final text.

\clearpage
\appendix
\section{Observer boundary terms}
\label{app:observer}

Proper time along the observer worldline obeys $\dd\tau=a(1+\Psi)\dd\eta$ at first order. Holding the observer's proper age fixed therefore shifts the Poisson-gauge coordinate time to $\eta_o=\bar\eta_o+\delta\eta_o$. Integrating from the initial hypersurface gives \cite{BiernYoo2017,YooScaccabarozzi2016}
\begin{equation}
 \delta\eta_o=\frac{a_{\rm in}}{a_o}\delta\eta_{\rm in}
       -\frac1{a_o}\int_{\eta_{\rm in}}^{\bar\eta_o}
                          \dd\eta\,a\Psi(\eta,\boldsymbol x_o).
 \label{eq:lapse}
\end{equation}
For regular initial conditions the first term vanishes. The lapse changes the coordinate length of the ray and the observer scale factor, so it enters both distance and redshift. It is a clock boundary term, separate from the observer velocity. The independent $\Phi_o$ term cancels in the observer tetrad \cite{BiernYoo2017,YooScaccabarozzi2016}.

The observer scalar contributions omitted from the main text are
\begin{align}
 \left.\frac{\delta D_A}{\bar D_A}\right|_{\rm obs,sc}
 &=\frac{\delta\eta_o}{r},
 \label{eq:observer-distance}\\
 \left.\frac{\delta z}{1+z}\right|_{\rm obs,sc}
 &=\Psi_o+\HH_o\delta\eta_o,
 \label{eq:observer-redshift}\\
 \left.\Delta_A\right|_{\rm obs,sc}
 &=-4\Q\Psi_o+
 \left[\frac{2-\beta_A}{r}+\frac1{r(1-r\HH)}-4\Q\HH_o\right]\delta\eta_o.
 \label{eq:observer-counts}
\end{align}
These are additive contributions to eqs.~\eqref{eq:distance-perturbation}, \eqref{eq:redshift-perturbation} and \eqref{eq:expanded-counts}, respectively. They depend on the observer event and background source distance, but not on direction, and hence contribute only to the monopole at first order. The observer velocity remains explicit in the main text. To restore the compact expression~\eqref{eq:countsA}, we add its direct volume term $2\delta\eta_o/r$ and use the full distance and redshift perturbations. All expressions in this appendix and appendix~\ref{app:general-selection} include the observer scalar terms.

With the clock term restored, the distance derivative is
\begin{equation}
 \frac{\dd}{\dd r}\left(\frac{\delta D_A}{\bar D_A}\right)
 =\Phi'-\partial_r\Phi+\frac{2\PhiW}{r}
 -\frac1{r^2}\int_0^r\dd\tilde r\,(2+\nabla^2_\Omega)\PhiW
 -\frac{\delta\eta_o}{r^2}.
 \label{eq:distance-derivative-expanded}
\end{equation}
The observer velocity is independent of source distance and drops out of this derivative. Substitution into the restored eq.~\eqref{eq:countsA} gives eq.~\eqref{eq:expanded-counts} plus the scalar contribution~\eqref{eq:observer-counts}.

The role of the observer terms can be seen by considering constant and pure-gradient adiabatic perturbations, which must not produce a count fluctuation. In an Einstein--de Sitter background, a constant potential $\Phi=\Psi=p$ has $V=-\eta p/3$, $\delta_g=0$, $\delta\eta_o=-\eta_op/3$ and $\eta_o=\eta+r$. Including eqs.~\eqref{eq:observer-distance} and \eqref{eq:observer-redshift}, the full perturbations are $\delta D_A/\bar D_A=p-\eta_op/(3r)$ and $\delta z/(1+z)=-2p/3$. Although these individual perturbations are non-zero, their contributions to the restored counts cancel, giving $\Delta_A=0$ for arbitrary $b_e$ and $\Q$.

For a pure gradient $\Phi=\Psi=\boldsymbol g\cdot\boldsymbol x$, the source and observer velocities are $\bv_s=-\eta\boldsymbol g/3$ and $\bv_o=-\eta_o\boldsymbol g/3$. At the observer origin, $\Psi_o=\delta\eta_o=0$. The full-sky convergence is $\kappa=-r\boldsymbol g\cdot\bn$, giving $\delta D_A/\bar D_A=(r-\eta_o/3)\boldsymbol g\cdot\bn$ and $\delta z/(1+z)=-2r\boldsymbol g\cdot\bn/3$. The count fluctuation again vanishes. The clock lapse, observer velocity and full-sky convergence are needed for these cancellations, together with the density-slicing conversion.

\Needspace{11\baselineskip}
\section{General radial observables and selection}
\label{app:general-selection}

Counts in any radial observable can be obtained by first calculating the counts at fixed emission time, then shifting to the time selected by the observed coordinate. For a radial observable $X$, with the observer fixed and perturbations evaluated at background emission time, the required shift is
\begin{equation}
 \delta\eta_X=-\frac{\delta X}{\dd\bar X/\dd\eta},
 \qquad
 \delta\eta_z=\frac1\HH\frac{\delta z}{1+z},
 \qquad
 \delta\eta_A=\frac r{1-r\HH}\frac{\delta D_A}{\bar D_A},
 \label{eq:general-time-displacements}
\end{equation}
while
\begin{equation}
 \delta\eta_L=\frac r{1+r\HH}
       \left(\frac{\delta D_A}{\bar D_A}
                        +2\frac{\delta z}{1+z}\right).
 \label{eq:general-time-displacement-L}
\end{equation}
The derivatives of the background quantities follow the background lightcone. The null condition gives
$\delta r_X=\delta\eta_o+2\int_0^r\PhiW\dd\tilde r-\delta\eta_X$.

Let $\Delta_\eta$ denote the gauge-dependent count fluctuation per coordinate emission time. Expanding the number current at the shifted event gives
\begin{align}
 \Delta_X&=\Delta_\eta+
       \left(\HH b_e-\frac2r\right)\delta\eta_X
       -\frac{\dd\delta\eta_X}{\dd r}
       +\Delta_X^{\rm sel},
 \label{eq:general-radial-template}\\
 \Delta_\eta&=\delta_n^{\rm P}+v_r-2\Phi+\Psi-2\kappa
       +\frac{2\delta\eta_o}{r}
       +\frac4r\int_0^r\dd\tilde r\,\PhiW
       +2\bv_o\cdot\bn.
 \label{eq:time-count-template}
\end{align}
Moving to a fixed observed coordinate changes the background density and transverse area through $\delta\eta_X$, and deforms the radial interval through its derivative. The shifts above, together with the density conversion in eq.~\eqref{eq:gauge}, give each count field without using a source Euler equation.

We describe a single intrinsic luminosity threshold by $L_{\rm cut}(1+z,D_A)$ and define its local slopes by
\begin{equation}
 p_z=\left.\frac{\partial\ln L_{\rm cut}}
                      {\partial\ln(1+z)}\right|_{D_A},
 \qquad
 p_A=\left.\frac{\partial\ln L_{\rm cut}}
                      {\partial\ln D_A}\right|_z.
 \label{eq:selection-slopes}
\end{equation}
The bias definitions in eq.~\eqref{eq:biases} continue to hold at fixed intrinsic threshold. At fixed radial coordinate,
\begin{align}
 \left.\delta\ln L_{\rm cut}\right|_X
 ={}&p_z\left(\frac{\delta z}{1+z}-\HH\delta\eta_X\right)
 \nonumber\\
 &+p_A\left[\frac{\delta D_A}{\bar D_A}
                  +\left(\HH-\frac1r\right)\delta\eta_X\right],
 \label{eq:general-threshold-perturbation}
\end{align}
and $\Delta_X^{\rm sel}=-\Q\,\delta\ln L_{\rm cut}|_X$. Holding $z$, $D_A$ or $D_L$ fixed in turn gives
\begin{align}
 \Delta_z^{\rm sel}
 &=-\Q p_A\left.\frac{\delta D_A}{\bar D_A}\right|_z,
 \nonumber\\
 \Delta_A^{\rm sel}
 &=-\Q p_z\left.\frac{\delta z}{1+z}\right|_{D_A},
 \nonumber\\
 \Delta_L^{\rm sel}
 &=-\Q(p_z-2p_A)\left.\frac{\delta z}{1+z}\right|_{D_L}.
 \label{eq:general-selection-terms}
\end{align}
The last line uses distance duality at fixed observed $D_L$.

The same threshold slopes determine how the selected background mean evolves:
\begin{equation}
 \frac{\dd\ln\bar n_{\rm sel}}{\dd\ln a}
 =b_e+\Q\left[p_z+p_A\left(\frac1{r\HH}-1\right)\right].
 \label{eq:general-selected-evolution}
\end{equation}
The total derivative follows the changing threshold along the selected population. Its difference from $b_e$, which holds that threshold fixed, is the selection contribution on the right-hand side.
The bolometric cut has $(p_z,p_A)=(4,2)$, recovering eqs.~\eqref{eq:selection} and \eqref{eq:selected-evolution}. Its luminosity-distance selection term vanishes because $D_L$ fixes the threshold.

For an angular-size threshold on an intrinsic physical radius, replacing $L_{\rm cut}$ by $R_{\rm cut}=\theta_{\rm cut}D_A$ and using the corresponding size slope gives $(p_z,p_A)=(0,1)$. Holding area distance fixed then holds the intrinsic radius threshold fixed. Population evolution and intrinsic size fluctuations still affect the counts.

\FloatBarrier
\section{The area-distance maximum and radial branches}
\label{app:branches}

\begin{figure}[t]
 \centering\includegraphics[width=.76\textwidth]{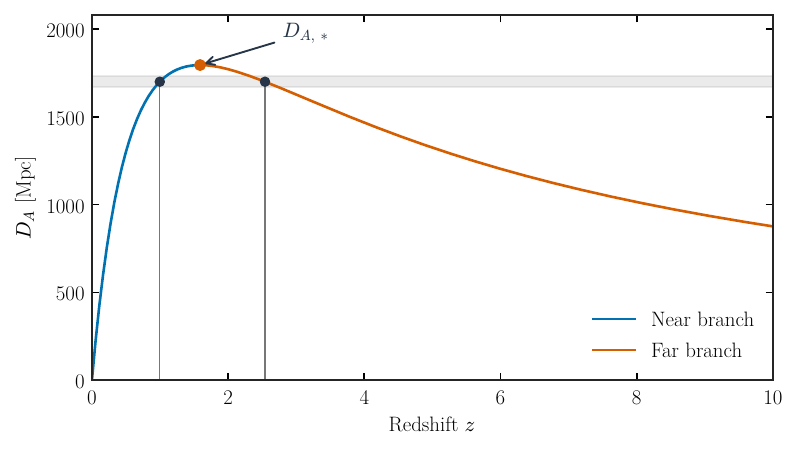}
 \caption{Area distance on the near and far radial branches in the flat matter-plus-$\Lambda$ model used for the numerical examples. The shaded distance interval selects two separated source intervals. The maximum is at $z_*=1.588$. A catalogue admits both roots only if its source population extends to them; an independent source label can select a single branch.}
 \label{fig:branches}
\end{figure}

The background area distance increases and then decreases along the lightcone, so one measured distance can correspond to two source shells (figure~\ref{fig:branches}). Near the maximum, a small distance interval covers a large radial interval. The count density per unit area distance consequently diverges, even though the number of sources in a finite interval remains finite.

If the same observed area distance occurs at several emission events on a ray, eq.~\eqref{eq:exact-radial-counts} gives
\begin{equation}
 N_A(\bn,D_A)=\sum_i\left.
       \frac{\rho(1+z)D_A^2}
            {|\dd D_A/\dd\lambda_{\rm aff}|}\right|_i,
 \label{eq:exact-optical-counts}
\end{equation}
The sum includes every event admitted by the catalogue. For the affinely parametrised $k^\mu$ in eq.~\eqref{eq:current}, the optical expansion is $\tfrac12\nabla_\mu k^\mu=\dd\ln D_A/\dd\lambda_{\rm aff}$. The denominator is therefore $\tfrac12D_A|\nabla_\mu k^\mu|$, so a region where the beam area varies slowly contributes more sources to a given distance interval.

At $r\HH=1$, the background distance derivative vanishes. The beam still has finite non-zero area; this is a degeneracy of the radial label, not an optical caustic. A catalogue selected only by area distance includes all admitted roots,
\begin{equation}
 \bar N_A(D_A)=\sum_{z_i:\bar D_A(z_i)=D_A}
       \frac{\bar N_z(z_i)}{|\dd\bar D_A/\dd z|_{z_i}}.
 \label{eq:branch-sum}
\end{equation}
Near a non-degenerate maximum, this density grows as $(D_{A,*}-D_A)^{-1/2}$. The divergence is integrable, but the pointwise perturbative expansion fails: a small change in distance can correspond to a large shift of the source label.

The single-branch expansion requires that this shift remain small and preserve the sign of the radial Jacobian. Close to a non-degenerate maximum, the leading local condition is
\begin{equation}
 D_{A,*}-\bar D_A\gg|\delta D_A|.
 \label{eq:fold-validity}
\end{equation}
If a distance perturbation is comparable to this gap, we can no longer expand the inverse distance relation about a single background root. We then calculate the number of sources in a finite bin directly from the current:
\begin{equation}
 \mathcal N_{[D_1,D_2]}(\bn)
 =\int_{\{\lambda_{\rm aff}:\,D_A\in[D_1,D_2]\}}
      \rho(1+z)D_A^2|\dd\lambda_{\rm aff}|.
 \label{eq:finite-bin-counts}
\end{equation}
This number per unit solid angle includes every admitted ray segment and stays finite across a smooth maximum. Bins containing the maximum must be integrated in a regular source coordinate before expansion.

For finite biases and $\HH_*'/\HH_*^2<1$, the pointwise area-distance dipole coefficient tends to negative infinity at the maximum. This marks the failure of the local radial expansion; angular projections, including section~\ref{sec:ellis-baldwin}, must use the full count distribution across the admitted branches. The local radial ruler interval in eq.~\eqref{eq:radial-ruler} also tends to zero there, so a finite ruler must be treated using its full radial extent.

\section{Area distance and angular diameter distance}
\label{app:angular-distance}

The usual cosmological angular diameter distance is defined by beam area and is the $D_A$ used here \cite{Fleury2015,Perlick2004}. To distinguish this definition from a distance inferred using a single diameter, we write the map from observed angular separations to source-screen separations as $\xi_s^a=\mathcal J^a{}_b\delta\theta_o^b$, where $a,b$ are screen indices. Its determinant gives
\begin{equation}
 D_A^2=\frac{\dd A_s}{\dd\Omega_o}=|\det\mathcal J|.
 \label{eq:angular-area-identity}
\end{equation}
This area relation holds in any spacetime, including for a sheared beam. Using the usual angular diameter distance therefore selects the same cells and gives the same count field $\Delta_A$.

A distance inferred from one diameter can depend on its orientation. For a unit direction $e^a$ on the observer's screen, the corresponding transverse ruler gives
\begin{equation}
 D_\theta(e)\equiv\frac{\dd\ell_s}{\dd\theta_o}
   =\bigl[e^a(\mathcal J^{\rm T}\mathcal J)_{ab}e^b\bigr]^{1/2}.
 \label{eq:directional-angular-distance}
\end{equation}
A small circular patch on the observer's sky maps to an ellipse on the source screen. If its angular radius is $\delta\theta_o$, the principal semiaxes are $s_1\delta\theta_o$ and $s_2\delta\theta_o$, where $s_1,s_2$ are the singular values of $\mathcal J$. A diameter along either principal direction therefore measures $s_1$ or $s_2$, while the area distance is their geometric mean, $D_A=\sqrt{s_1s_2}$ \cite{Perlick2004}.

For a fixed diameter convention and unchanged population, count conservation on a locally invertible segment gives
\begin{equation}
 N_\theta(\bn,D_\theta)
 =N_A(\bn,D_A)\left|\frac{\dd D_A}{\dd D_\theta}\right|.
 \label{eq:angular-count-jacobian}
\end{equation}
The beam area in eq.~\eqref{eq:exact-radial-counts} remains $D_A^2\dd\Omega_o$. Write $D_\theta=D_A(1+\epsilon)$, with $\epsilon$ first order and $\bar D_\theta=\bar D_A$. Expanding eq.~\eqref{eq:angular-count-jacobian} about the same background distance gives
\begin{align}
 \Delta_\theta-\Delta_A
 &=-\frac1{\bar N_A}\frac{\dd}{\dd\bar D_A}
          \bigl(\bar N_A\bar D_A\epsilon\bigr)
 \nonumber\\
 &=-\frac r{1-r\HH}\frac{\dd\epsilon}{\dd r}
      -(\beta_A-2\Q)\epsilon.
 \label{eq:angular-count-correction}
\end{align}
The first line follows from count conservation. The bolometric selection gives the second through $1+\dd\ln\bar N_A/\dd\ln\bar D_A=\beta_A-2\Q$.

For linear scalar perturbations, write $\mathcal J/\bar D_A=(1+\delta D_A/\bar D_A)I-\gamma$, where $\gamma_{ab}$ is the trace-free image shear. Equation~\eqref{eq:directional-angular-distance} gives
\begin{equation}
 \epsilon(e)\equiv\frac{D_\theta(e)-D_A}{D_A}
 =-\gamma_{ab}e^ae^b=-\gamma\cos2\varphi
 \qquad\hbox{at first order}.
 \label{eq:directional-shear-correction}
\end{equation}
Here the shear eigenvalues are $+\gamma$ and $-\gamma$, with $\gamma\geq0$, and $\varphi$ is the angle between the chosen diameter and the $+\gamma$ principal axis. The fractional distance difference $\epsilon$ is therefore the shear projected along that diameter: it is $-\gamma$ and $+\gamma$ on the two principal axes and averages to zero for uniformly weighted orientations. An area-equivalent size has $\epsilon=0$ exactly. Additional size or shape cuts change the population and require the corresponding selection response, as in appendix~\ref{app:general-selection}.

\FloatBarrier
\section{Numerical angular spectra}
\label{app:numerical-spectra}

\subsection{Fourier range and numerical inputs}

For the numerical examples, we use $\Omega_{m0}=0.3153$, $H_0=67.36\,{\rm km\,s^{-1}\,Mpc^{-1}}$ and scale-independent growing perturbations, neglecting radiation. The matter spectrum is the smooth Eisenstein--Hu spectrum \cite{EisensteinHu1998}, with $\Omega_{b0}=0.0493$, $n_s=0.9649$, $\sigma_8=0.8111$ and $T_{\rm CMB}=2.7255\,{\rm K}$. The count spectra use $b=1$, $b_e=1$ and $\Q=0.5$. The spectrum has no baryon acoustic oscillations and is normalised to $\sigma_8$ with a spherical top-hat of radius $8h^{-1}\,{\rm Mpc}$, where $h=0.6736$. 

We integrate from $k_{\min}=10^{-6}\,{\rm Mpc}^{-1}$ without a Limber approximation or radial averaging. The upper Fourier cutoff is $k_{\max}=0.16\,{\rm Mpc}^{-1}$. We multiply $P_0(k)$ by a common cosine-squared taper, equal to unity for $k\leq0.12\,{\rm Mpc}^{-1}$, falling smoothly to zero at $k_{\max}$ and vanishing above it.

At a fixed radial position, an angular multipole receives contributions from radial wavenumbers much larger than $\ell/r$. The plotted spectra and quoted percentages therefore depend on the Fourier cutoff. At low redshift the spectra turn down when the cutoff removes modes contributing to the angular power.

We stop the $z=0.3$ curves in figures~\ref{fig:spectra}, \ref{fig:space-differences}, \ref{fig:component-autopowers}, \ref{fig:cross-contributions} and \ref{fig:coordinate-lensing-appendix} at $\ell=60$, before this downturn; the other redshifts extend to $\ell=120$. The selection and fixed-area-distance redshift decompositions retain their calculated range through $\ell=120$, showing the cutoff dependence at low redshift. A prediction for a radially selected catalogue must also include its radial window and distance errors.

\subsection{Components and their correlations}

\begin{figure}[t]
 \centering\includegraphics[width=\textwidth]{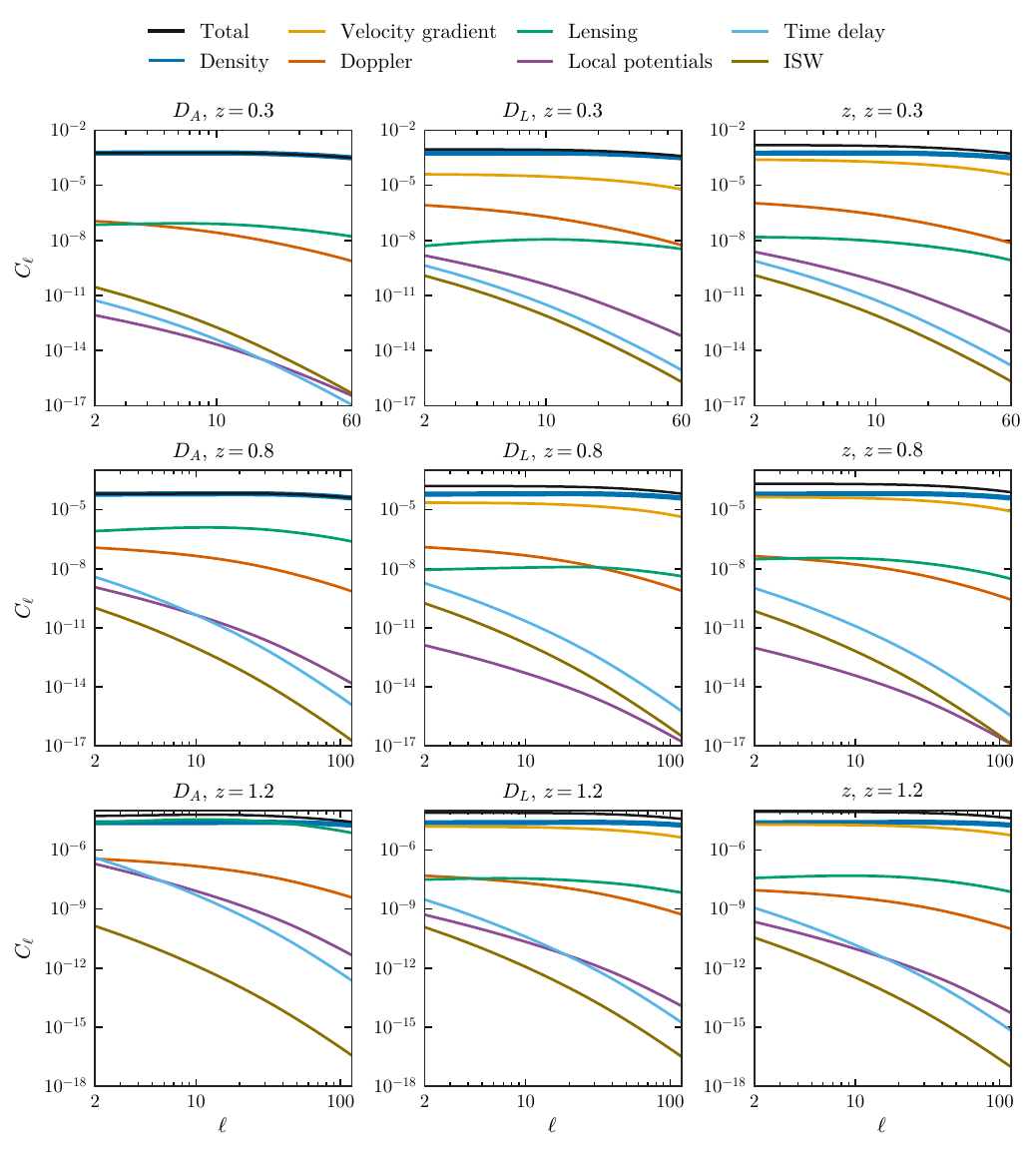}
 \caption{Individual contributions to the count auto-spectra $C_\ell^{AA}$, $C_\ell^{LL}$ and $C_\ell^{zz}$, arranged by radial coordinate (columns) and background source redshift (rows). Each coloured curve is the auto-power of one term in the count field. The black curve is the full auto-spectrum, including every cross term, so it is not the sum of the coloured curves. The velocity-gradient contribution vanishes in area distance space. Local potentials include the density-slicing conversion; time delay and ISW denote the remaining integrals of $\PhiW$ and $\PhiW'$, respectively. Lensing includes both $\kappa$ and its radial derivative where present. The total and density curves nearly coincide in area distance space at low redshift. The population and Fourier selection are the same as in figure~\ref{fig:spectra}, with $\ell\leq60$ at $z=0.3$ and $\ell\leq120$ otherwise.}
 \label{fig:component-autopowers}
\end{figure}

Figure~\ref{fig:component-autopowers} separates the auto-power of each contribution, following ref.~\cite{Fonseca2023}. We group the count terms by their physical origin: the matter-comoving tracer density, the velocity gradient responsible for the Kaiser distortion, and the local Doppler terms with the growing-mode coefficients in eqs.~\eqref{eq:velocityA}--\eqref{eq:velocityz}. The lensing contribution includes both the convergence and its radial derivative, since both arise from the same change in beam area. 
The remaining terms are divided into local potentials, including the density-slicing term $(3-b_e)\HH V$, the time-delay integral of $\PhiW$, and the integrated Sachs--Wolfe (ISW) integral of $\PhiW'$. Observer terms contribute only at $\ell=0,1$ and are absent from these plots.

\begin{figure}[t]
 \centering\includegraphics[width=\textwidth]{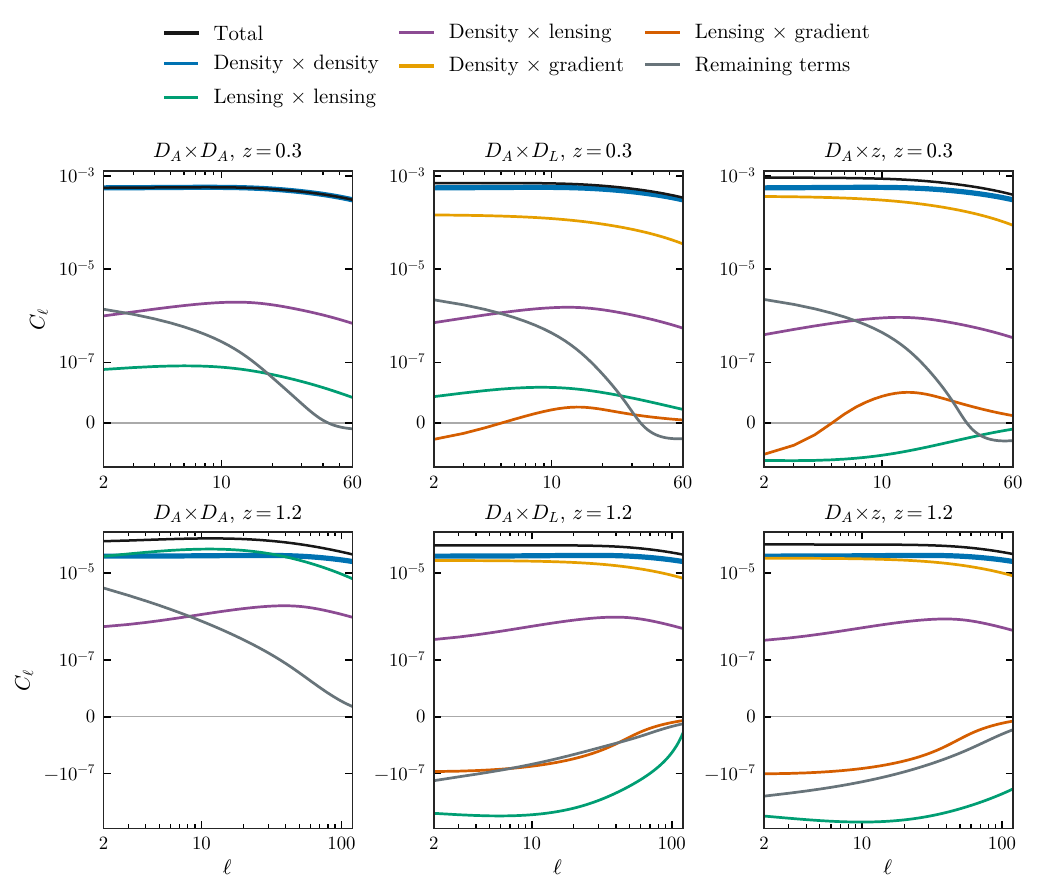}
 \caption{Auto- and cross-term contributions to $C_\ell^{AA}$, $C_\ell^{AL}$ and $C_\ell^{Az}$ at $z=0.3$ (top) and $z=1.2$ (bottom). Mixed entries include both orderings of the two named contributions. The remaining terms contain all auto- and cross-correlations involving Doppler, local potentials, time delay or ISW. The coloured curves therefore add to the black total. Terms that vanish identically are omitted. The vertical axes use an asinh scale, retaining negative contributions and zero crossings; all panels in a row use the same scale. The source positions, biases and Fourier selection match figure~\ref{fig:spectra}.}
 \label{fig:cross-contributions}
\end{figure}

Correlations between the count terms also contribute to the full spectrum. Figure~\ref{fig:cross-contributions} shows these correlations explicitly for the density, velocity-gradient and lensing terms in $C_\ell^{AA}$, $C_\ell^{AL}$ and $C_\ell^{Az}$. Each mixed entry includes both orderings of its two contributions, giving the usual factor of two in an auto-spectrum. Since $\Delta_A$ has no velocity-gradient term, the gradient contributions appear only in its cross-spectra with luminosity distance and redshift. They correlate with both density and lensing, and can either raise or lower the cross-power.

\begin{figure}[t]
 \centering\includegraphics[width=\textwidth]{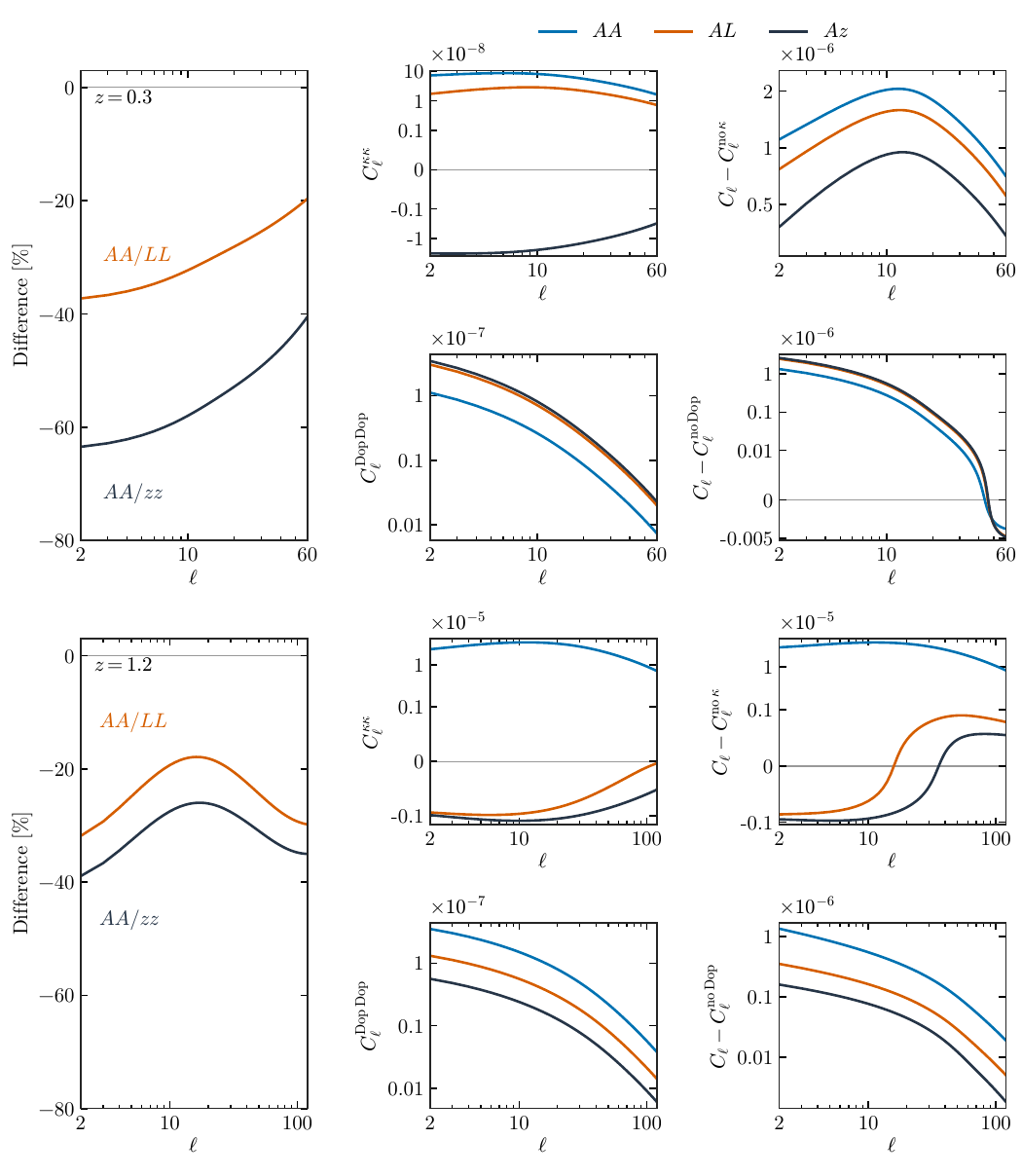}
 \caption{The coordinate, lensing and Doppler comparison of figure~\ref{fig:space-differences} at $z=0.3$ (upper pair of rows) and $1.2$ (lower pair). For each redshift, the left panel shows the relative area-distance auto-power; the middle and right panels show the isolated-term spectra and the change on removing them, with lensing above and Doppler below. Colours and definitions follow figure~\ref{fig:space-differences}. The $z=0.3$ curves stop at $\ell=60$, before the downturn induced by the Fourier cutoff used to limit nonlinear modes; the $z=1.2$ curves extend to $\ell=120$.}
 \label{fig:coordinate-lensing-appendix}
\end{figure}

Removing a contribution from the count field changes its auto-power and all its cross-correlations with the remaining terms. Figures~\ref{fig:space-differences} and \ref{fig:coordinate-lensing-appendix} show this change for local Doppler terms and convergence. The local Doppler terms are the $v_r$ terms in eqs.~\eqref{eq:velocityA}--\eqref{eq:velocityz}, where $\partial_r\Psi=-(3\Omega_m/2f)\HH v_r$ has already been combined with the explicit velocities. Kaiser gradients remain in the rest of the field. Removing the Doppler terms changes the spectrum by
\begin{equation}
 C_\ell^{XY}-C_\ell^{XY,\mathrm{no\ Dop}}
 =C_\ell^{XY,\mathrm{Dop\,Dop}}
 +C_\ell^{XY,\mathrm{Dop\,rest}}
 +C_\ell^{XY,\mathrm{rest\,Dop}}.
 \label{eq:component-spectrum}
\end{equation}
Here ``rest'' contains every other count term. Cross-correlations with these terms can make the full difference larger than the Doppler auto-power alone. The separate Doppler and convergence differences share their mutual cross terms; adding the two differences would count those correlations twice.

\Needspace{8\baselineskip}
\subsection{Evolution and flux selection}

\begin{figure}[tp]
 \centering\includegraphics[width=.93\textwidth]{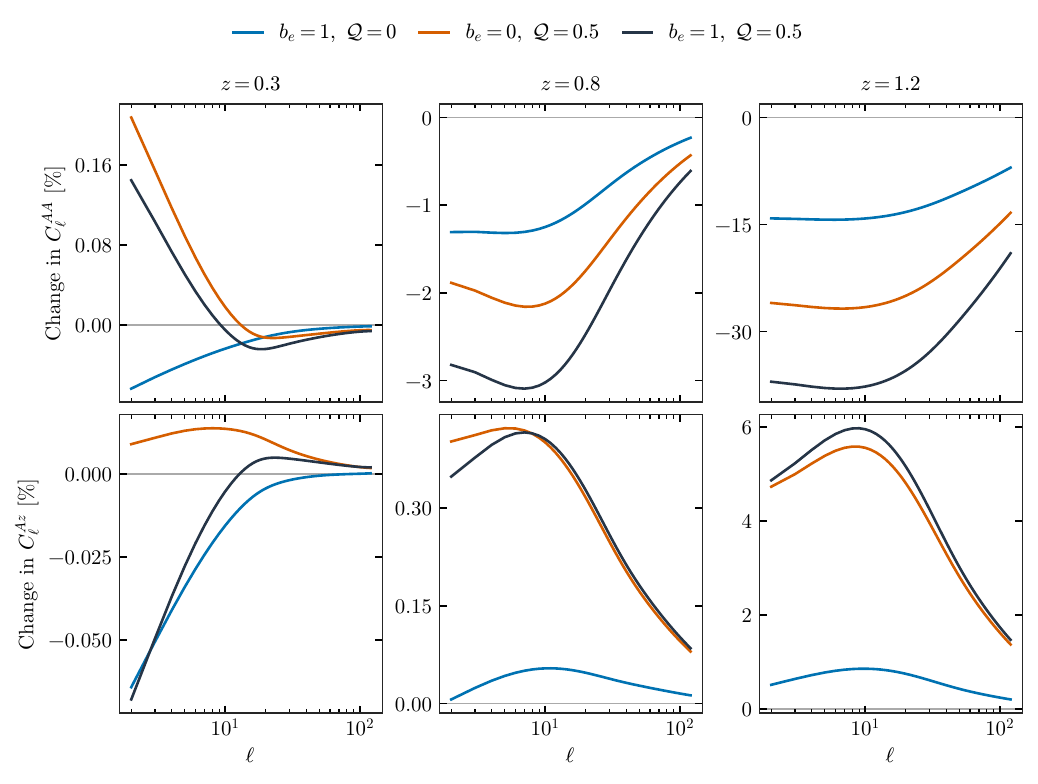}
 \caption{Effects of evolution and flux selection on $C_\ell^{AA}$ (top) and $C_\ell^{Az}$ (bottom). Curves show percentage changes from $b_e=\Q=0$, varying evolution alone, flux selection alone, and both. The matter spectrum, density bias, source positions and Fourier cutoff are held fixed.}
 \label{fig:bias-effects}
\end{figure}

Figure~\ref{fig:bias-effects} separates the effects of $b_e$ and $\Q$. Evolution changes the density assigned to the shifted emission time; flux selection changes which sources enter the catalogue. For $b_e=1$, $\Q=0.5$, their combined effect lowers $C_2^{AA}$ at $z=1.2$ by $37\%$ relative to the zero-bias case. The values are illustrative; applying the result to a survey requires the biases of its selected population.

\FloatBarrier
\Needspace{8\baselineskip}
\subsection{Redshift fluctuations at fixed area distance}
\label{app:fixed-distance-spectrum}

The density-cancelling combination in eq.~\eqref{eq:count-identity} is proportional to the fractional redshift perturbation at fixed area distance. To calculate the spectrum of this perturbation, we combine the redshift and distance transfer functions as
\begin{equation}
 T_\ell^{z|D_A}
 =T_\ell^{\delta z/(1+z)}
  -\frac{r\HH}{1-r\HH}T_\ell^{\delta D_A/\bar D_A}.
 \label{eq:redshift-transfer}
\end{equation}
We obtain its auto-spectrum from eq.~\eqref{eq:angular-spectra} by using this transfer for both fields. At $\ell\geq2$, this field consists of the source radial velocity and an optical contribution,
\begin{equation}
 -\Psi-2\int_0^r\dd\tilde r\,\PhiW'
 -\frac{r\HH}{1-r\HH}
  \left[-\Phi+\frac2r\int_0^r\dd\tilde r\,\PhiW-\kappa\right].
 \label{eq:optical-redshift}
\end{equation}
This term contains gravitational redshift, integrated potentials and lensing. Observer scalars and velocity project onto $\ell=0$ and $1$, respectively, so they remain in the general fields but not in these plots.

\begin{figure}[tp]
 \centering\includegraphics[width=.93\textwidth]{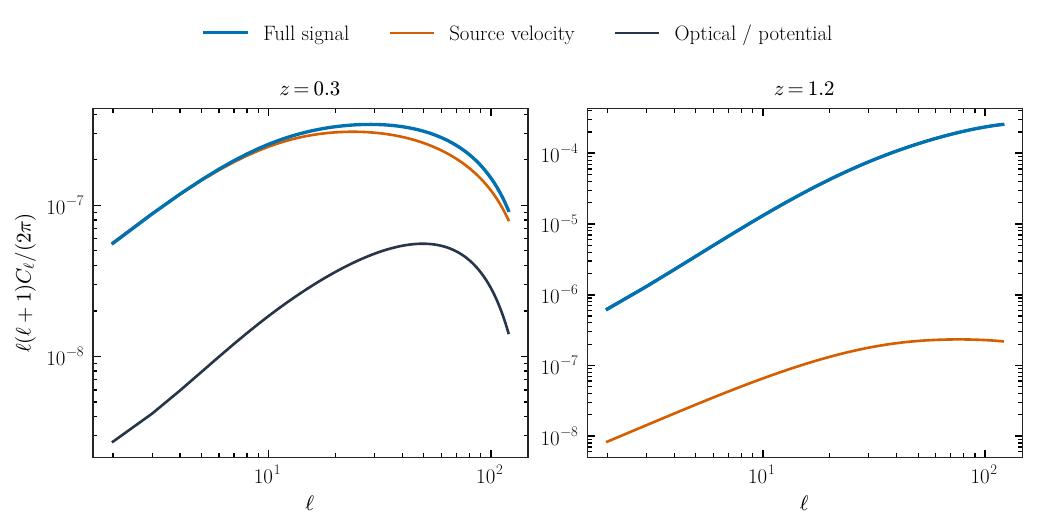}
 \caption{Angular power of $\delta z|_{D_A}/(1+z)$. Blue: full signal; orange: source-velocity auto-power; dark curve: optical and gravitational auto-power from eq.~\eqref{eq:optical-redshift}. Their cross term is included in the full signal. The full and optical curves overlap at $z=1.2$. This field is independent of $b$, $b_e$ and $\Q$.}
 \label{fig:redshift-spectrum}
\end{figure}

At $z=0.3$, the velocity auto-power in figure~\ref{fig:redshift-spectrum} exceeds the optical auto-power by factors of $21$ at $\ell=2$ and $6$ at $\ell=120$. Optical terms dominate by $z=0.8$. Their growth with distance is amplified by the area-distance radial Jacobian. Unlike the individual count spectra, this combination is independent of the population biases.

\FloatBarrier


\begin{thebibliography}{99}

\bibitem{Kaiser1987}
N.~Kaiser, \href{https://doi.org/10.1093/mnras/227.1.1}{\emph{Clustering in real space and in redshift space}}, Mon. Not. R. Astron. Soc. \textbf{227} (1987) 1--21.

\bibitem{Hamilton1998}
A.~J.~S.~Hamilton, \href{https://doi.org/10.1007/978-94-011-4960-0_17}{\emph{Linear redshift distortions: a review}}, The Evolving Universe (Astrophysics and Space Science Library) \textbf{231} (1998) 185--275, \href{https://arxiv.org/abs/astro-ph/9708102}{arXiv:astro-ph/9708102}.

\bibitem{Yoo2009}
J.~Yoo, A.~L.~Fitzpatrick and M.~Zaldarriaga, \href{https://doi.org/10.1103/PhysRevD.80.083514}{\emph{New perspective on galaxy clustering as a cosmological probe: general relativistic effects}}, Phys. Rev. D \textbf{80} (2009) 083514, \href{https://arxiv.org/abs/0907.0707}{arXiv:0907.0707}.

\bibitem{Bonvin2011}
C.~Bonvin and R.~Durrer, \href{https://doi.org/10.1103/PhysRevD.84.063505}{\emph{What galaxy surveys really measure}}, Phys. Rev. D \textbf{84} (2011) 063505, \href{https://arxiv.org/abs/1105.5280}{arXiv:1105.5280}.

\bibitem{Challinor2011}
A.~Challinor and A.~Lewis, \href{https://doi.org/10.1103/PhysRevD.84.043516}{\emph{Linear power spectrum of observed source number counts}}, Phys. Rev. D \textbf{84} (2011) 043516, \href{https://arxiv.org/abs/1105.5292}{arXiv:1105.5292}.

\bibitem{Jeong2012}
D.~Jeong, F.~Schmidt and C.~M.~Hirata, \href{https://doi.org/10.1103/PhysRevD.85.023504}{\emph{Large-scale clustering of galaxies in general relativity}}, Phys. Rev. D \textbf{85} (2012) 023504, \href{https://arxiv.org/abs/1107.5427}{arXiv:1107.5427}.

\bibitem{Bertacca2012}
D.~Bertacca, R.~Maartens, A.~Raccanelli and C.~Clarkson, \href{https://doi.org/10.1088/1475-7516/2012/10/025}{\emph{Beyond the plane-parallel and Newtonian approach: wide-angle redshift distortions and convergence in general relativity}}, JCAP \textbf{10} (2012) 025, \href{https://arxiv.org/abs/1205.5221}{arXiv:1205.5221}.

\bibitem{Tansella2018}
V.~Tansella, C.~Bonvin, R.~Durrer, B.~Ghosh and E.~Sellentin, \href{https://doi.org/10.1088/1475-7516/2018/03/019}{\emph{The full-sky relativistic correlation function and power spectrum of galaxy number counts. Part I: theoretical aspects}}, JCAP \textbf{03} (2018) 019, \href{https://arxiv.org/abs/1708.00492}{arXiv:1708.00492}.

\bibitem{Castorina2022}
E.~Castorina and E.~di Dio, \href{https://doi.org/10.1088/1475-7516/2022/01/061}{\emph{The observed galaxy power spectrum in general relativity}}, JCAP \textbf{01} (2022) 061, \href{https://arxiv.org/abs/2106.08857}{arXiv:2106.08857}.

\bibitem{Alonso2015}
D.~Alonso, P.~Bull, P.~G.~Ferreira, R.~Maartens and M.~G.~Santos, \href{https://doi.org/10.1088/0004-637X/814/2/145}{\emph{Ultra large-scale cosmology in next-generation experiments with single tracers}}, Astrophys. J. \textbf{814} (2015) 145, \href{https://arxiv.org/abs/1505.07596}{arXiv:1505.07596}.

\bibitem{Namikawa2016}
T.~Namikawa, A.~Nishizawa and A.~Taruya, \href{https://doi.org/10.1103/PhysRevLett.116.121302}{\emph{Anisotropies of gravitational-wave standard sirens as a new cosmological probe without redshift information}}, Phys. Rev. Lett. \textbf{116} (2016) 121302, \href{https://arxiv.org/abs/1511.04638}{arXiv:1511.04638}.

\bibitem{Zhang2018}
P.~Zhang, \href{https://arxiv.org/abs/1810.11915}{\emph{The large scale structure in the 3D luminosity-distance space and its cosmological applications}}, (2018), \href{https://arxiv.org/abs/1810.11915}{arXiv:1810.11915}.

\bibitem{Namikawa2021}
T.~Namikawa, \href{https://doi.org/10.1088/1475-7516/2021/01/036}{\emph{Analyzing clustering of astrophysical gravitational-wave sources: luminosity-distance space distortions}}, JCAP \textbf{01} (2021) 036, \href{https://arxiv.org/abs/2007.04359}{arXiv:2007.04359}.

\bibitem{Libanore2021}
S.~Libanore et al., \href{https://doi.org/10.1088/1475-7516/2021/02/035}{\emph{Gravitational wave mergers as tracers of large scale structures}}, JCAP \textbf{02} (2021) 035, \href{https://arxiv.org/abs/2007.06905}{arXiv:2007.06905}.

\bibitem{Libanore2022}
S.~Libanore et al., \href{https://doi.org/10.1088/1475-7516/2022/02/003}{\emph{Clustering of gravitational wave and supernovae events: a multitracer analysis in luminosity distance space}}, JCAP \textbf{02} (2022) 003, \href{https://arxiv.org/abs/2109.10857}{arXiv:2109.10857}.

\bibitem{Fonseca2023}
J.~Fonseca, S.~Zazzera, T.~Baker and C.~Clarkson, \href{https://doi.org/10.1088/1475-7516/2023/08/050}{\emph{The observed number counts in luminosity distance space}}, JCAP \textbf{08} (2023) 050, \href{https://arxiv.org/abs/2304.14253}{arXiv:2304.14253}.

\bibitem{FonsecaErratum2024}
J.~Fonseca, S.~Zazzera, T.~Baker and C.~Clarkson, \href{https://doi.org/10.1088/1475-7516/2024/06/E02}{\emph{Erratum: The observed number counts in luminosity distance space}}, JCAP \textbf{06} (2024) E02.

\bibitem{Balaudo2024}
A.~Balaudo, M.~Pantiri and A.~Silvestri, \href{https://doi.org/10.1088/1475-7516/2024/02/023}{\emph{Number count of gravitational waves and supernovae in luminosity distance space for {$\Lambda$CDM} and scalar-tensor theories}}, JCAP \textbf{02} (2024) 023, \href{https://arxiv.org/abs/2311.17904}{arXiv:2311.17904}.

\bibitem{Zazzera2025}
S. Zazzera, J. Fonseca, T. Baker and C. Clarkson, \href{https://arxiv.org/abs/2412.01678}{\emph{Gravitational waves and galaxies cross-correlations: a forecast on GW biases for future detectors}}, Mon. Not. R. Astron. Soc. \textbf{537} (2025) 1912--1923, \href{https://arxiv.org/abs/2412.01678}{arXiv:2412.01678}.

\bibitem{Zazzera2026}
S. Zazzera, J. Fonseca, T. Baker and C. Clarkson, \href{https://arxiv.org/abs/2505.15645}{\emph{Exploring future synergies for large-scale structure between gravitational waves and radio sources}}, Mon. Not. R. Astron. Soc. \textbf{547} (2026) stag307, \href{https://arxiv.org/abs/2505.15645}{arXiv:2505.15645}.

\bibitem{KristianSachs1966}
J.~Kristian and R.~K.~Sachs, \href{https://doi.org/10.1086/148522}{\emph{Observations in cosmology}}, Astrophys. J. \textbf{143} (1966) 379--399.

\bibitem{Fleury2015}
P.~Fleury, \href{https://arxiv.org/abs/1511.03702}{\emph{Light propagation in inhomogeneous and anisotropic cosmologies}}, (2015), PhD thesis, Universit\'e Pierre et Marie Curie, \href{https://arxiv.org/abs/1511.03702}{arXiv:1511.03702}.

\bibitem{Schmidt2012}
F.~Schmidt and D.~Jeong, \href{https://doi.org/10.1103/PhysRevD.86.083527}{\emph{Cosmic rulers}}, Phys. Rev. D \textbf{86} (2012) 083527, \href{https://arxiv.org/abs/1204.3625}{arXiv:1204.3625}.

\bibitem{Etherington1933}
I.~M.~H.~Etherington, \href{https://doi.org/10.1080/14786443309462220}{\emph{On the definition of distance in general relativity}}, Philos. Mag. \textbf{15} (1933) 761--773.

\bibitem{Sachs1961}
R.~K.~Sachs, \href{https://doi.org/10.1098/rspa.1961.0202}{\emph{Gravitational waves in general relativity. VI. The outgoing radiation condition}}, Proc. R. Soc. Lond. A \textbf{264} (1961) 309--338.

\bibitem{Alsing2015}
J.~Alsing, D.~Kirk, A.~Heavens and A.~H.~Jaffe, \href{https://doi.org/10.1093/mnras/stv1249}{\emph{Weak lensing with sizes, magnitudes and shapes}}, Mon. Not. R. Astron. Soc. \textbf{452} (2015) 1202--1216, \href{https://arxiv.org/abs/1410.7839}{arXiv:1410.7839}.

\bibitem{Ye2026}
I.~Ye, P.~Bull, C.~Guandalin, C.~Clarkson and A.~Nasirudin, \href{https://doi.org/10.1093/mnras/stag506}{\emph{Prospects for measuring the Doppler magnification dipole with LSST and DESI}}, Mon. Not. R. Astron. Soc. \textbf{547} (2026) stag506, \href{https://arxiv.org/abs/2511.17858}{arXiv:2511.17858}.

\bibitem{Bonvin2008}
C.~Bonvin, \href{https://doi.org/10.1103/PhysRevD.78.123530}{\emph{Effect of peculiar motion in weak lensing}}, Phys. Rev. D \textbf{78} (2008) 123530, \href{https://arxiv.org/abs/0810.0180}{arXiv:0810.0180}.

\bibitem{Bacon2014}
D.~J.~Bacon, S.~Andrianomena, C.~Clarkson, K.~Bolejko and R.~Maartens, \href{https://doi.org/10.1093/mnras/stu1270}{\emph{Cosmology with Doppler lensing}}, Mon. Not. R. Astron. Soc. \textbf{443} (2014) 1900--1915, \href{https://arxiv.org/abs/1401.3694}{arXiv:1401.3694}.

\bibitem{BonvinDipole2017}
C.~Bonvin, S.~Andrianomena, D.~Bacon, C.~Clarkson, R.~Maartens, T.~Moloi and P.~Bull, \href{https://doi.org/10.1093/mnras/stx2049}{\emph{Dipolar modulation in the size of galaxies: the effect of Doppler magnification}}, Mon. Not. R. Astron. Soc. \textbf{472} (2017) 3936--3951, \href{https://arxiv.org/abs/1610.05946}{arXiv:1610.05946}.

\bibitem{Andrianomena2019}
S.~Andrianomena, C.~Bonvin, D.~Bacon, P.~Bull, C.~Clarkson, R.~Maartens and T.~Moloi, \href{https://doi.org/10.1093/mnras/stz1905}{\emph{Testing general relativity with the Doppler magnification effect}}, Mon. Not. R. Astron. Soc. \textbf{488} (2019) 3759--3771, \href{https://arxiv.org/abs/1810.12793}{arXiv:1810.12793}.

\bibitem{MaartensBias2021}
R.~Maartens, J.~Fonseca, S.~Camera, S.~Jolicoeur, J.-A.~Viljoen and C.~Clarkson, \href{https://doi.org/10.1088/1475-7516/2021/12/009}{\emph{Magnification and evolution biases in large-scale structure surveys}}, JCAP \textbf{12} (2021) 009, \href{https://arxiv.org/abs/2107.13401}{arXiv:2107.13401}.

\bibitem{Jain2011}
B.~Jain and M.~Lima, \href{https://doi.org/10.1111/j.1365-2966.2010.17505.x}{\emph{Magnification effects on source counts and fluxes}}, Mon. Not. R. Astron. Soc. \textbf{411} (2011) 2113--2117, \href{https://arxiv.org/abs/1003.6127}{arXiv:1003.6127}.

\bibitem{Duncan2022}
C.~A.~J.~Duncan, J.~Harnois-D{\'e}raps, L.~Miller and A.~Langedijk, \href{https://doi.org/10.1093/mnras/stac1809}{\emph{On cosmological bias due to the magnification of shear and position samples in modern weak lensing analyses}}, Mon. Not. R. Astron. Soc. \textbf{515} (2022) 1130--1145, \href{https://arxiv.org/abs/2111.09867}{arXiv:2111.09867}.

\bibitem{Zazzera2024}
S.~Zazzera, J.~Fonseca, T.~Baker and C.~Clarkson, \href{https://doi.org/10.1088/1475-7516/2024/05/095}{\emph{Magnification and evolution bias of transient sources: {GWs} and {SNIa}}}, JCAP \textbf{05} (2024) 095, \href{https://arxiv.org/abs/2309.04391}{arXiv:2309.04391}.

\bibitem{Ellis1971}
G.~F.~R.~Ellis, \href{https://doi.org/10.1007/s10714-009-0760-7}{\emph{Relativistic cosmology}}, in \emph{General Relativity and Cosmology}, Proceedings of the International School of Physics ``Enrico Fermi'', Course 47, R.~K.~Sachs, ed., Academic Press (1971), pp.~104--182; republished in \href{https://doi.org/10.1007/s10714-009-0760-7}{Gen. Relativ. Gravit. \textbf{41} (2009) 581--660}.

\bibitem{Ellis1985}
G.~F.~R.~Ellis, S.~D.~Nel, R.~Maartens, W.~R.~Stoeger and A.~P.~Whitman, \href{https://doi.org/10.1016/0370-1573(85)90030-4}{\emph{Ideal observational cosmology}}, Phys. Rep. \textbf{124} (1985) 315--417.

\bibitem{Bonvin2006}
C. Bonvin, R. Durrer and M. A. Gasparini, \href{https://arxiv.org/abs/astro-ph/0511183}{\emph{Fluctuations of the luminosity distance}}, Phys. Rev. D \textbf{73} (2006) 023523; erratum, Phys. Rev. D \textbf{85} (2012) 029901, \href{https://arxiv.org/abs/astro-ph/0511183}{arXiv:astro-ph/0511183}.

\bibitem{HuiGreene2006}
L. Hui and P. B. Greene, \href{https://arxiv.org/abs/astro-ph/0512159}{\emph{Correlated fluctuations in luminosity distance and the importance of peculiar motion in supernova surveys}}, Phys. Rev. D \textbf{73} (2006) 123526, \href{https://arxiv.org/abs/astro-ph/0512159}{arXiv:astro-ph/0512159}.

\bibitem{Bonvin2014}
C. Bonvin, L. Hui and E. Gazta\~naga, \href{https://arxiv.org/abs/1309.1321}{\emph{Asymmetric galaxy correlation functions}}, Phys. Rev. D \textbf{89} (2014) 083535, \href{https://arxiv.org/abs/1309.1321}{arXiv:1309.1321}.

\bibitem{BonvinFleury2018}
C.~Bonvin and P.~Fleury, \href{https://arxiv.org/abs/1803.02771}{\emph{Testing the equivalence principle on cosmological scales}}, JCAP \textbf{05} (2018) 061, \href{https://arxiv.org/abs/1803.02771}{arXiv:1803.02771}.

\bibitem{PaulClarkson2026}
P.~Paul and C.~Clarkson, \href{https://doi.org/10.1088/1475-7516/2026/06/044}{\emph{Visualising relativistic effects in redshift space distortions of large scale structure}}, JCAP \textbf{06} (2026) 044, \href{https://arxiv.org/abs/2508.03249}{arXiv:2508.03249}.

\bibitem{BonvinDistanceDipole2006}
C.~Bonvin, R.~Durrer and M.~Kunz, \href{https://doi.org/10.1103/PhysRevLett.96.191302}{\emph{Dipole of the luminosity distance: a direct measure of $H(z)$}}, Phys. Rev. Lett. \textbf{96} (2006) 191302, \href{https://arxiv.org/abs/astro-ph/0603240}{arXiv:astro-ph/0603240}.

\bibitem{MaartensDipole2018}
R.~Maartens, C.~Clarkson and S.~Chen, \href{https://doi.org/10.1088/1475-7516/2018/01/013}{\emph{The kinematic dipole in galaxy redshift surveys}}, JCAP \textbf{01} (2018) 013, \href{https://arxiv.org/abs/1709.04165}{arXiv:1709.04165}.

\bibitem{Kalbouneh2024}
B.~Kalbouneh, C.~Marinoni and R.~Maartens, \href{https://arxiv.org/abs/2401.12291}{\emph{Cosmography of the local Universe by multipole analysis of the expansion rate fluctuation field}}, JCAP \textbf{09} (2024) 069, \href{https://arxiv.org/abs/2401.12291}{arXiv:2401.12291}.

\bibitem{Kalbouneh2026}
B.~Kalbouneh, C.~Marinoni, R.~Maartens, J.~Bel, J.~Santiago, C.~Clarkson, M.~Sarma and J.-M.~Virey, \href{https://doi.org/10.1088/1475-7516/2026/03/084}{\emph{The anisotropic expansion rate of the local Universe and its covariant cosmographic interpretation}}, JCAP \textbf{03} (2026) 084, \href{https://arxiv.org/abs/2510.02510}{arXiv:2510.02510}.

\bibitem{EllisBaldwin1984}
G.~F.~R.~Ellis and J.~E.~Baldwin, \href{https://doi.org/10.1093/mnras/206.2.377}{\emph{On the expected anisotropy of radio source counts}}, Mon. Not. R. Astron. Soc. \textbf{206} (1984) 377--381.

\bibitem{vonHausegger2024}
S.~von Hausegger, \href{https://doi.org/10.1093/mnrasl/slae092}{\emph{The expected kinematic matter dipole is robust against source evolution}}, Mon. Not. R. Astron. Soc. Lett. \textbf{535} (2024) L49--L53, \href{https://arxiv.org/abs/2404.07929}{arXiv:2404.07929}.

\bibitem{vonHauseggerDalang2025}
S.~von Hausegger and C.~Dalang, \href{https://doi.org/10.1103/PhysRevD.111.123547}{\emph{Redshift tomography of the kinematic matter dipole}}, Phys. Rev. D \textbf{111} (2025) 123547, \href{https://arxiv.org/abs/2412.13162}{arXiv:2412.13162}.

\bibitem{Bonnefous2026}
A.~Bonnefous, \href{https://doi.org/10.1051/0004-6361/202659275}{\emph{The kinematic cosmic dipole beyond Ellis and Baldwin}}, Astron. Astrophys. \textbf{708} (2026) A357, \href{https://arxiv.org/abs/2602.05700}{arXiv:2602.05700}.

\bibitem{Huff2014}
E.~M.~Huff and G.~J.~Graves, \href{https://doi.org/10.1088/2041-8205/780/2/L16}{\emph{Magnificent magnification: exploiting the other half of the lensing signal}}, Astrophys. J. Lett. \textbf{780} (2014) L16, \href{https://arxiv.org/abs/1111.1070}{arXiv:1111.1070}.

\bibitem{SchmidtMagnification2012}
F.~Schmidt, A.~Leauthaud, R.~Massey, J.~Rhodes, M.~R.~George, A.~M.~Koekemoer, A.~Finoguenov and M.~Tanaka, \href{https://doi.org/10.1088/2041-8205/744/2/L22}{\emph{A detection of weak lensing magnification using galaxy sizes and magnitudes}}, Astrophys. J. Lett. \textbf{744} (2012) L22, \href{https://arxiv.org/abs/1111.3679}{arXiv:1111.3679}.

\bibitem{Umeh2014I}
O. Umeh, C. Clarkson and R. Maartens, \href{https://arxiv.org/abs/1207.2109}{\emph{Nonlinear relativistic corrections to cosmological distances, redshift and gravitational lensing magnification. I. Key results}}, Class. Quantum Grav. \textbf{31} (2014) 202001, \href{https://arxiv.org/abs/1207.2109}{arXiv:1207.2109}.

\bibitem{Umeh2014II}
O. Umeh, C. Clarkson and R. Maartens, \href{https://arxiv.org/abs/1402.1933}{\emph{Nonlinear relativistic corrections to cosmological distances, redshift and gravitational lensing magnification. II. Derivation}}, Class. Quantum Grav. \textbf{31} (2014) 205001, \href{https://arxiv.org/abs/1402.1933}{arXiv:1402.1933}.

\bibitem{Bolejko2013}
K.~Bolejko, C.~Clarkson, R.~Maartens, D.~Bacon, N.~Meures and E.~Beynon, \href{https://doi.org/10.1103/PhysRevLett.110.021302}{\emph{Antilensing: the bright side of voids}}, Phys. Rev. Lett. \textbf{110} (2013) 021302, \href{https://arxiv.org/abs/1209.3142}{arXiv:1209.3142}.

\bibitem{Joachimi2015}
B.~Joachimi, S.~Singh and R.~Mandelbaum, \href{https://doi.org/10.1093/mnras/stv1962}{\emph{Detection of spatial correlations of Fundamental Plane residuals, and cosmological implications}}, Mon. Not. R. Astron. Soc. \textbf{454} (2015) 478--488, \href{https://arxiv.org/abs/1504.02662}{arXiv:1504.02662}.

\bibitem{Ciarlariello2015}
S.~Ciarlariello, R.~Crittenden and F.~Pace, \href{https://doi.org/10.1093/mnras/stv447}{\emph{Intrinsic size correlations in weak lensing}}, Mon. Not. R. Astron. Soc. \textbf{449} (2015) 2059--2068, \href{https://arxiv.org/abs/1412.4606}{arXiv:1412.4606}.

\bibitem{Ciarlariello2016}
S.~Ciarlariello and R.~Crittenden, \href{https://doi.org/10.1093/mnras/stw2052}{\emph{Modelling the impact of intrinsic size and luminosity correlations on magnification estimation}}, Mon. Not. R. Astron. Soc. \textbf{463} (2016) 740--755, \href{https://arxiv.org/abs/1607.08784}{arXiv:1607.08784}.

\bibitem{Paul2023}
P. Paul, C. Clarkson and R. Maartens, \href{https://arxiv.org/abs/2208.04819}{\emph{Wide-angle effects in multi-tracer power spectra with Doppler corrections}}, JCAP \textbf{04} (2023) 067, \href{https://arxiv.org/abs/2208.04819}{arXiv:2208.04819}.

\bibitem{Alcock1979}
C.~Alcock and B.~Paczy{\'n}ski, \href{https://doi.org/10.1038/281358a0}{\emph{An evolution free test for non-zero cosmological constant}}, Nature \textbf{281} (1979) 358--359.

\bibitem{Ballinger1996}
W.~E.~Ballinger, J.~A.~Peacock and A.~F.~Heavens, \href{https://doi.org/10.1093/mnras/282.3.877}{\emph{Measuring the cosmological constant with redshift surveys}}, Mon. Not. R. Astron. Soc. \textbf{282} (1996) 877--888, \href{https://arxiv.org/abs/astro-ph/9605017}{arXiv:astro-ph/9605017}.

\bibitem{Seo2003}
H.-J.~Seo and D.~J.~Eisenstein, \href{https://doi.org/10.1086/379122}{\emph{Probing dark energy with baryonic acoustic oscillations from future large galaxy redshift surveys}}, Astrophys. J. \textbf{598} (2003) 720--740, \href{https://arxiv.org/abs/astro-ph/0307460}{arXiv:astro-ph/0307460}.

\bibitem{AlonsoDM2021}
D.~Alonso, \href{https://doi.org/10.1103/PhysRevD.103.123544}{\emph{Linear anisotropies in dispersion-measure-based cosmological observables}}, Phys. Rev. D \textbf{103} (2021) 123544, \href{https://arxiv.org/abs/2103.14016}{arXiv:2103.14016}.

\bibitem{BiernYoo2017}
S.~G.~Biern and J.~Yoo, \href{https://doi.org/10.1088/1475-7516/2017/04/045}{\emph{Gauge-invariance and infrared divergences in the luminosity distance}}, JCAP \textbf{04} (2017) 045, \href{https://arxiv.org/abs/1606.01910}{arXiv:1606.01910}.

\bibitem{YooScaccabarozzi2016}
J.~Yoo and F.~Scaccabarozzi, \href{https://doi.org/10.1088/1475-7516/2016/09/046}{\emph{Unified treatment of the luminosity distance in cosmology}}, JCAP \textbf{09} (2016) 046, \href{https://arxiv.org/abs/1606.08453}{arXiv:1606.08453}.

\bibitem{Perlick2004}
V.~Perlick, \href{https://doi.org/10.12942/lrr-2004-9}{\emph{Gravitational lensing from a spacetime perspective}}, Living Rev. Relativ. \textbf{7} (2004) 9, \href{https://arxiv.org/abs/1010.3416}{arXiv:1010.3416}.

\bibitem{EisensteinHu1998}
D.~J.~Eisenstein and W.~Hu, \href{https://doi.org/10.1086/305424}{\emph{Baryonic features in the matter transfer function}}, Astrophys. J. \textbf{496} (1998) 605--614, \href{https://arxiv.org/abs/astro-ph/9709112}{arXiv:astro-ph/9709112}.

\end{thebibliography}
\end{document}